\documentclass[aps,pre,reprint,floatfix,nofootinbib]{revtex4-2}
\usepackage[utf8]{inputenc}
\usepackage{siunitx}
\usepackage{graphicx}
\usepackage{physics}
\usepackage{amsmath}
\usepackage{amssymb}
\usepackage{bm}
\usepackage{dsfont}
\usepackage[ruled,vlined,linesnumbered]{algorithm2e}
\usepackage[dvipsnames]{xcolor}
\usepackage[normalem]{ulem}

\newcommand{\dir}{\bm{\hat{\Omega}}}
\newcommand{\pos}{\bm{r}}
\newcommand{\pt}{\pos,\dir,E}
\newcommand{\Et}{\Sigma_t}
\newcommand{\Es}{\Sigma_s}
\newcommand{\Ef}{\Sigma_f}
\newcommand{\keff}{k_{\text{eff}}}
\newcommand{\dirto}[2]{\frac{#2 - #1}{\abs{#2 - #1}}}
\newcommand{\nubar}{\bar{\nu}}
\newcommand{\RR}{\mathcal{R}}
\newcommand{\fs}{f_\text{scat}}
\newcommand{\ff}{f_\text{fiss}}
\newcommand{\fbar}{\bar{f}}

\begin{document}
    \title{Unbiasedness and Optimization of Regional Weight Cancellation}
    
    \author{Hunter Belanger}
    \email{hunter.belanger@cea.fr}
    \author{Davide Mancusi}
    \email{davide.mancusi@cea.fr}
    \author{Andrea Zoia}
    \email{andrea.zoia@cea.fr}
    
    \affiliation{Universit\'e Paris-Saclay, CEA, Service d'\'Etudes des R\'eacteurs et de Math\'ematiques Appliqu\'ees, 91191, Gif-sur-Yvette, France}
    
    \begin{abstract}
        The Monte Carlo method is often used to simulate systems which can be modeled by random walks. In order to calculate observables, in many implementations the ``walkers'' carry a statistical weight which is generally assumed to be positive. Some random walk simulations, however, may require walkers to have positive or negative weights: it has been shown that the presence of a mixture of positive and negative weights can impede the statistical convergence, and special weight-cancellation techniques must be adopted in order to overcome these issues. In a recent work we demonstrated the usefulness of one such method, exact regional weight cancellation, to solve eigenvalue problems in nuclear reactor physics in three spatial dimensions. The method previously exhibited had several limitations (including multi-group transport and isotropic scattering) and needed homogeneous cuboid cancellation regions. In this paper we lift the previous limitations, in view of applying exact regional cancellation to more realistic continuous-energy neutron transport problems. This extended regional cancellation framework is used to optimize the efficiency of the weight cancellation. Our findings are illustrated on a benchmark configuration for reactor physics.
    \end{abstract}
    
    \maketitle

    \section{Introduction}

        For day-to-day industrial needs in the field of nuclear reactor physics, deterministic codes are used to solve the neutron transport equation, estimating the reaction rates and the power distribution in the reactor core \cite{AP3,Collins2016,CASMO5}. Deterministic methods have the advantage of running fast, but this speed comes at the cost of accuracy: many approximations are introduced, discretizing the phase space (position, direction and energy) and thus leading to a bias in the results. The gold-standard in reactor physics for solving the neutron transport equation is the Monte Carlo method, which does not need to introduce any discretization of the phase space, and is therefore free of any bias \cite{the_sacred_text}. This high-fidelity simulation method comes at the cost of requiring extensive computer resources. Because of this computational cost, multi-physics simulations of a full-core nuclear reactor model, using Monte Carlo neutronics codes coupled with other state-of-the-art thermal-hydraulics and fuel performance codes, have become possible only very recently, mainly thanks to the increase in available computer power and to the development of efficient variance-reduction techniques \cite{Romano2020, Mancusi2022}.
        
        In these Monte Carlo simulations, the particles being simulated (typically neutrons or photons) carry a statistical weight, which is used to estimate observable quantities such as reaction rates and power distributions within the core of the nuclear reactor. For most applications involved in nuclear reactor physics or radiation shielding problems, these statistical weights are always positive. However, several types of Monte Carlo neutronics simulations require that the particles being sampled carry negative statistical weights (or complex weights, where each component is allowed to be negative). Problems that require negative weights include the evaluation of the second harmonic of the flux, critical buckling, and neutron noise, as the quantities being estimated in these problems can be negative \cite{Yamamoto2012,Booth2003,Yamamoto2012,Rouchon2017}. There are also special rejection sampling methods which allow negative weights, that could be used to treat spatially continuous material properties, even when the desired quantities should be positive \cite{Belanger2020}. Random walk problems using positive and negative statistical weights emerge more broadly in many applications outside of nuclear reactor physics, e.g.\ in quantum diffusion Monte Carlo \cite{Arnow1982}, or in the Wigner Monte Carlo formalism \cite{Sellier2015}. Such simulations can be particularly challenging, as the summing of positive and negative contributions to estimate the observable quantities leads to very large variances in these tallies: it is often recognized that weight cancellation is mandatory to ensure convergence \cite{Booth2003,Yamamoto2012,Assaraf2007}. 
        
        In a recent work, we have focused on the case of spatially continuous material properties for particle transport applications emerging in reactor physics. Material cross sections for neutron transport depend on the energy of the incident particle, as well as on the temperature and density of the material. Traditional neutronics codes (both Monte Carlo and deterministic) make the approximation that each material region in the reactor model has a constant temperature and density \cite{T4,OpenMC,Serpent,MCNP,AP3,Collins2016,CASMO5}. In a real nuclear reactor, however, this is certainly not the case, as the temperature and density will depend continuously on position. The continual advances of high-performance computing resources allows us to consider new ways of improving the fidelity of our Monte Carlo codes. It is in this context that we have examined the possibility of treating spatially-continuous material temperatures and densities in Monte Carlo simulations in a previous work \cite{Belanger2020}. In particular, we have focused on assessing which particle-tracking methods might be best suitable to treat spatially-continuous cross sections for fixed-source transport problems, typically occurring in radiation shielding applications \cite{Belanger2020}. Among the possible choices, the delta-tracking \cite{Woodcock1965,Leppanen2017} and negative-weighted delta-tracking \cite{Carter1972, Legrady2017} sampling strategies were deemed particularly attractive. Negative-weighted delta-tracking, although beneficial for dealing with spatially-continuous cross sections, has a potential drawback due to the statistical weights of the particles being allowed to become negative: in a subsequent study concerning $k$-eigenvalue problems, we have shown that the coupling of positive and negative particle weights prevents convergence of the power iteration method to the fundamental mode of the physical system being studied \cite{Belanger2021}.

        To overcome these issues, an exact regional weight cancellation method, originally proposed by Booth and Gubernatis in a 1D context \cite{Booth2010}, was extended to 3D and was shown to allow the convergence of power iteration with negative-weighted delta-tracking in a multi-group reactor physics benchmark \cite{Belanger2021}. Such a weight cancellation technique might be useful to improve the simulation methods of the other previously mentioned problems which have particles with negative statistical weights. While our previous work in Ref.~\citenum{Belanger2021} demonstrated potential for the method, many questions were left unanswered: under what conditions is regional cancellation unbiased? How might one extend cancellation from multi-group to continuous-energy material cross sections? Is it possible to maximize the efficiency of weight cancellation, for a given set of particles in a cancellation region? Our goal in this paper is to build upon our previous results in Ref.~\citenum{Belanger2021} and to start addressing these very questions.
        
        Our manuscript is organised as follows. In Sec.~\ref{sec:summary_previous_paper}, we provide a brief summary of the exact regional cancellation technique which we presented in our previous paper. Section~\ref{sec:unbiasedness} will develop the mathematical theory behind the family of techniques for regional cancellation, for the most general continuous-energy case. We also examine what conditions must be satisfied to ensure that a regional cancellation method is unbiased. The theory presented in this section elucidates the concepts which are integral to the technique (for both multi-group and continuous-energy calculations), and blazes the trail for implementing exact regional cancellation in continuous-energy problems. The question of optimizing cancellation is then treated in Sec.~\ref{sec:optimization}, where two candidate optimization methods are proposed. Section~\ref{sec:implementation} discusses the implementation of the two optimization strategies in our Monte Carlo code, and discusses how these two strategies allow us to deal with heterogeneous cancellation regions. The different optimization strategies are compared in Sec.~\ref{sec:results}, and we also assess the performances of our methods on a reactor physics benchmark with heterogeneous cancellation regions. Some concluding thoughts and remarks are provided in Sec.~\ref{sec:conclusions}.

    \section{Regional Weight Cancellation}\label{sec:summary_previous_paper}
    
        Previously, we have extended the 1D exact regional cancellation scheme of Booth and Gubernatis \cite{Booth2010} to work in 3D multi-group neutron transport problems \cite{Belanger2021}. Here, we shall briefly outline the mechanics of this method, in a general continuous-energy framework. For the case of $k$-eigenvalue problems, the fundamental mode and eigenvalue are sought by Monte Carlo methods using power iteration, which basically consists of following the neutron histories over fission generations \cite{the_sacred_text}. When negative-weighted delta-tracking is used to sample particle flights, the transported neutron will have positive and negative weights, and weight cancellation will be mandatory to ensure the convergence of power iteration \cite{Belanger2021}. In this context, the regional cancellation operation is applied to neutrons born from fission. The fission particles are first sorted into user-defined cancellation regions, based on their position. Once all of the particles have been sorted into their cancellation regions, we may then consider each cancellation region independently for the cancellation procedure. In our previous work, a simple rectilinear mesh was imposed on top of the problem geometry.
        
        Consider cancellation region $\RR$ (which is assumed to be composed of only one fissile material), containing fission neutrons which have already been sampled. In addition to storing its own position ($\pos$), energy ($E$), and direction ($\dir$), each fission particle also stores its parent's energy ($E'$), the position of the previous collision ($\pos'$), and the direction of the parent's penultimate flight ($\dir''$).\footnote{The direction of the parent's last flight is not explicitly stored, as it can be calculated as $\dir'=\dirto{\pos'}{\pos}$.} From this information, we can calculate the ``fission density function'', i.e.\ the expected fission density at $\pos$ due to a collision at $\pos'$ coming from direction $\dir''$, and a subsequent flight from $\pos'$ to $\pos$ at energy $E'$; this is a key ingredient for the weight cancellation procedure. The exact form of the fission density function depends on the particle tracking method being used. For the case of negative-weighted delta-tracking,\footnote{We will only mention negative-weighted delta-tracking in the text, since that was the focus of our previous work, but Eq.~\eqref{eq:previous_paper_zeta} is also valid for regular delta-tracking, where $\Sigma_\text{smp}$ would be the majorant cross section. This could be of use for neutron noise or critical buckling problems, which would not necessarily require the use of negative-weighted delta-tracking but nonetheless require weight cancellation.} as examined in our previous work, the fission density function was taken to be
        \begin{multline}
            \zeta(\pos|\pos',\dir'',E') = \\
            \frac{\mathcal{P}\left(\dirto{\pos'}{\pos}\cdot\dir''\right)
                  \Ef(\pos,E')}
                 {2\pi\abs{\pos-\pos'}^2}
            e^{-\Sigma_\text{smp}(E')\abs{\pos-\pos'}}
            \text.
            \label{eq:previous_paper_zeta}
        \end{multline}
        In this notation, $\Sigma_\text{smp}$ is the sampling cross-section required for negative-weighted delta tracking, $\Ef$ is the fission cross-section, and $\mathcal{P}$ is the probability density function for the cosine of the scattering angle for the previous collision.\footnote{While the symbol $f$ was used for the fission density function in Ref.~\citenum{Belanger2021}, we have instead chosen to use $\zeta$ in this paper, to avoid any confusion  with other subsequent symbols.} Based on $\zeta$, we are able to split each fission particle in $\RR$ into two components: a point-wise component with weight $w_p$, and a uniform component of weight $w_u$. The point-wise portion, $w_p$, keeps the phase space coordinates $(\pos,\dir,E)$ of the split fission particle. The uniform component, $w_u$, is spread uniformly over the region $\RR$. To calculate the point-wise and uniform weights, we use
        \begin{align}
            w_p = \frac{\zeta(\pos|\pos',\dir'',E') - \beta}{\zeta(\pos|\pos',\dir'',E')}w
            \label{eq:w_p} \\
            w_u = \frac{\beta}{\zeta(\pos|\pos',\dir'',E')}w \label{eq:w_u}
            \text,
        \end{align}
        respectively, with $w$ being the weight of the original fission particle \cite{Booth2010}. Note that $w_p+w_u=w$, so that the net weight is conserved. The free parameter $\beta$ can take any value, and in general is chosen independently for each particle in $\RR$. Our previous work followed the recommendation of Booth and Gubernatis, and always took $\beta$ to be the minimum value of $\zeta(\pos''|\pos',\dir'',E')$ over all possible $\pos'' \in \RR$, for the particle of interest. We demonstrated that, for the case of isotropic scattering and cuboid cancellation regions, one only needs to evaluate $\zeta(\pos''|\pos',\dir'',E')$ for the eight corners of the cuboid (taking $\pos''$ to be the corner positions) to find the minimum value within $\RR$.
        
        With $w_p$ and $w_u$ having been calculated for each fission particle in $\RR$, we then take the sum of all the uniform weight components
        \begin{equation}
            U = \sum_{i=1}^N w_{u,i}
            \text,
        \end{equation}
        where the extra subscript $i$ indicates the fission particle.
        This operation is effectively where the cancellation occurs: depending on the initial weights $w_i$ of the fission particles, the individual uniform components $w_{u,i}$ will be positive or negative, and taking their sum cancels some of the positive and negative weight which was in the region $\RR$. The uniform weight $U$ must be distributed uniformly within $\RR$. To do this,
        $n = \lceil \abs{U} \rceil$
        new fission particles are sampled within $\RR$, each having a weight of $U/n$.
        The positions of the $n$ uniform particles are sampled uniformly in $\RR$. In our previous work, the direction was sampled from an isotropic distribution, as fission was assumed to be perfectly isotropic, and the energy was sampled from the fission spectrum of the material in $\RR$, as it was assumed that the fission spectrum had no dependence on incident neutron energy. These $n$ new uniform fission particles must be added to the fission bank, and will then be transported along with the other fission particles during the next fission generation.
        
        The method proposed in Ref.~\citenum{Belanger2021} that we have recalled here, was demonstrated to work successfully and be unbiased on a simple reactor physics benchmark problem. While those results were very promising, the initial implementation admittedly had several limitations. First, cancellation regions must be homogeneous, containing only a single fissile material. Second, fission must always be isotropic, and the fission spectrum must be independent of the incident energy. In general, even in continuous-energy transport, fission is almost always represented as isotropic, so this is not necessarily a large inconvenience. However, the fission energy spectrum is generally assumed to be dependent on the incident neutron energy. Furthermore, while Booth and Gubernatis argue that cancellation is unbiased for any value of the parameter $\beta$, the amount of canceled weight (and thus the efficiency of the method) clearly does depend on $\beta$. Using the minimum value of the fission density as $\beta$ is not necessarily the most efficient choice for achieving the highest amount of cancellation. Nonetheless, taking $\beta$ to be the minimum within $\RR$ guarantees that both $w_p$ and $w_u$ have the same sign as $w$: when $\beta$ is larger than the minimum, the point-wise portion, $w_p$, can change sign, potentially leading to even more positive and negative weight in the region than there was initially. The cancellation operation does not change the net weight $W_\text{net}$ in the bin, as
        \begin{equation}
            W_\text{net} = \sum_i^N w_i = \sum_i^N w_{u,i} + \sum_i^N w_{p,i} = U + \sum_i^N w_{p,i}
            \label{eq:W_net}
        \end{equation}
        will still be located in the bin. However, cancellation does change the total weight, $W_\text{tot}$, defined as the sum of the absolute values of all weights.
        The total weight before cancellation is
        \begin{equation}
            W_\text{tot} = \sum_i^N \abs{w_i}
            \text,
            \label{eq:W_tot}
        \end{equation}
        while the post-cancellation total weight is
        \begin{equation}
            W_\text{tot,post} = \sum_i^N \abs{w_{p,i}} + \abs{U} = \sum_i^N \abs{w_{p,i}} + \abs{\sum_i^N w_{u,i}}
            \text.
            \label{eq:post_cancellation_total_weight}
        \end{equation}
        By using the triangle inequality, it is possible to show that
        \[
        W_\text{tot,post}\geq |W_\text{net}|
        \text.
        \]
        The more efficient cancellation is, the closer $W_\text{tot,post}$ will be to $W_\text{net}$, with 100\% cancellation efficiency corresponding to $W_\text{tot,post} = W_\text{net}$ (i.e.\ all negative weight is removed). The optimal choice for $\beta$ will maximize the cancellation efficiency, and therefore minimize $W_\text{tot,post}$. This optimal choice of $\beta$ is clearly dependent on the other particles in the bin, and determining this optimal value is vital for improving the overall computational efficiency of the simulation.
    
    \section{Unbiasedness of Cancellation}\label{sec:unbiasedness}
        
        In this section, a method for performing exact regional cancellation in general continuous-energy problems shall be developed, and it will be demonstrated under what conditions such schemes lead to an unbiased fission source. For this purpose, it is mathematically beneficial to use the integral form of the transport equation, as opposed to the integro-differential form adopted in our previous work. We will begin by presenting the integral transport form for the eigenvalue transport problem in Sec.~\ref{sec:integral_eqs}. Section~\ref{sec:estimators_averaging_over_all} makes a first attempt at developing an estimator for the fission emission density in a region, which averages over all possible collisions and subsequent flights which induce the fission. While this exact estimator is likely of little use to a practical application, we are able to use it to examine what requirements must be observed in order to have an unbiased fission emission density estimator. Section~\ref{sec:expected_value_estimators} discusses how far back in a particle's history one must look, so that it is possible for it to have contributed to the fission emission density everywhere within the cancellation region. Section~\ref{sec:condition_on_sampled_scatter} uses the ideas from Section~\ref{sec:expected_value_estimators} to decompose the collision operator as is done in most Monte Carlo codes, to produce a fission emission density estimator which could potentially be used in an industrial code to achieve exact regional cancellation. Section~\ref{sec:non-uniform} outlines the possibility of distributing some of the fission emission density within the region according to a generic function, instead of distributing it uniformly. Finally, Section~\ref{sec:cancellation_and_dt} examines why delta-tracking algorithms are more suited to exact regional cancellation, and the peculiarities which can arise from delta-scatters.

        \subsection{Integral Formulation of the Transport Equation}
        \label{sec:integral_eqs}
        
        We will start with the $k$-eigenvalue Boltzmann transport equation in integral form. Let $P=(\pt)$ denote the coordinates of a point in phase space. The collision density $\psi(P)=\Et(\pos,E)\varphi(\pt)$ and the emission density $\chi(P)$ are related by \cite{the_sacred_text}:
        \begin{align}
          \psi(P)&{}=\mathbb{T}\chi(P)
          \label{eq:psi}\\
          \chi(P)&{}=\left[\mathbb{C}_s + \frac{1}{k}\mathbb{C}_f\right]\psi(P) \text,
          \label{eq:chi}
        \end{align}
        where $\Et$ is the total macroscopic cross section, and $\varphi$ is the angular neutron flux.
        In this notation, $\mathbb{T}$ is the flight operator,
        defined as
        \begin{equation}
            \mathbb{T}g(P) =
            \int T(P'\to P)g(P')\dd P'
            \text,
            \label{eq:flight_op}
        \end{equation}
        where we have made use of the flight kernel
        \begin{multline}
            T(P'\to P) = \\
            \frac{\Et(\pos,E')}
                 {\abs{\pos-\pos'}^2}
            \exp\left(
                      -\int\limits_0^{\abs{\pos-\pos'}}\Et(\pos' + u\dir',E')\dd u
                \right) \\
            \delta\left(\dir - \dirto{\pos'}{\pos}\right)
            \delta\left(\dir'-\dir\right)
            \delta\left(E-E'\right)
            \text.
            \label{eq:flight_kernel}
        \end{multline}
        We note that the flight kernel $T(P'\to P)$ is normalized, and can be interpreted as the probability density function (PDF) for a particle having a flight and landing at the phase space coordinate $P$, conditioning on its initial phase space coordinate being $P'$. 
        
        The scattering operator $\mathbb{C}_s$ in Eq.~\eqref{eq:chi} is defined as
        \begin{equation}
            \mathbb{C}_sg(P) =
            \int C_s(P'\to P)g(P')\dd P'
            \text,
            \label{eq:scat_op}
        \end{equation}
        with the scattering kernel $C_s(P'\to P)$ being
        \begin{multline}
            C_s(P'\to P) =
            \frac{\nu_s(\pos',E')\Es(\pos',E')}
                 {\Et(\pos',E')} \\
            \fs\left(\dir,E|\pos',\dir',E'\right)
            \delta\left(\pos-\pos'\right)
            \text,
            \label{eq:col_scat}
        \end{multline}
        where $\Es$ is the macroscopic scattering cross section, $\nu_s$ is the average number of neutrons emitted from a scatter, and $\fs$ is the joint PDF for a neutron to scatter in direction $\dir$ at energy $E$.
        The fission operator $\mathbb{C}_f$ is similar to the scattering operator in Eq.~\eqref{eq:scat_op}, but instead uses a fission kernel $C_f(P'\to P)$, defined as
        \begin{multline}
            C_f(P'\to P) =
            \frac{\nu_f(\pos',E')\Ef(\pos',E')}
                 {\Et(\pos',E')} \\
            \ff\left(\dir,E|\pos',\dir',E'\right)
            \delta\left(\pos-\pos'\right)
            \text,
            \label{eq:col_fiss}
        \end{multline}
        where $\Ef$ is the macroscopic fission cross section, $\nu_f$ is the average number of neutrons produced per fission, and $\ff$ is the joint PDF for fission neutrons to be emitted in direction $\dir$ at energy $E$.
        The scattering and fission operators may be combined into a collision operator
        \begin{equation}
            \chi(P) = \mathbb{C}\psi(P)
            \text,
        \end{equation}
        which has a corresponding collision kernel
        \begin{equation}
           C(P'\to P) = C_s(P'\to P) + \frac{1}{k} C_f(P'\to P)
           \text.
        \end{equation}
        Here $C(P'\to P)$ can be interpreted as the average number of of particles produced about the phase space coordinate $P$, from a collision induced by a particle at $P'$. Given this interpretation, it is also possible to rewrite $C(P'\to P)$ in a more concise form, using an average yield $\nubar(\pos',E')$, and an average transfer function $\fbar(\dir,E | \pos', \dir', E')$:
        \begin{equation}
            C(P'\to P) =
            \nubar(\pos',E')
            \fbar\left(\dir,E | \pos', \dir', E'\right)
            \delta(\pos - \pos')
            \text.
            \label{eq:C_kernel_simp}
        \end{equation}
        It is clear that Eq.~\eqref{eq:C_kernel_simp} is true if
        \begin{equation}
            \nubar(\pos',E') =
            \frac{\nu_s(\pos',E')\Es(\pos',E')}
                 {\Et(\pos',E')} +
            \frac{\nu_f(\pos',E')\Ef(\pos',E')}
                 {k\Et(\pos',E')}
        \end{equation}
        and
        \begin{multline}
            \fbar\left(\dir,E | \pos', \dir', E'\right) = \\
            \frac{\nu_s(\pos',E')\Es(\pos',E')}
                 {\nubar(\pos',E')\Et(\pos',E')}
            \fs\left(\dir,E|\pos',\dir',E'\right) + \\
            \frac{\nu_f(\pos',E')\Ef(\pos',E')}
                 {k\nubar(\pos',E')\Et(\pos',E')}
            \ff\left(\dir,E|\pos',\dir',E'\right)
            \text.
            \label{eq:fbar}
        \end{multline}

        \subsection{Averaging over all Scattering Events}
        \label{sec:estimators_averaging_over_all}
        
        \begin{figure*}
            \begin{minipage}[c]{0.483\textwidth}
            \includegraphics[width=\textwidth]{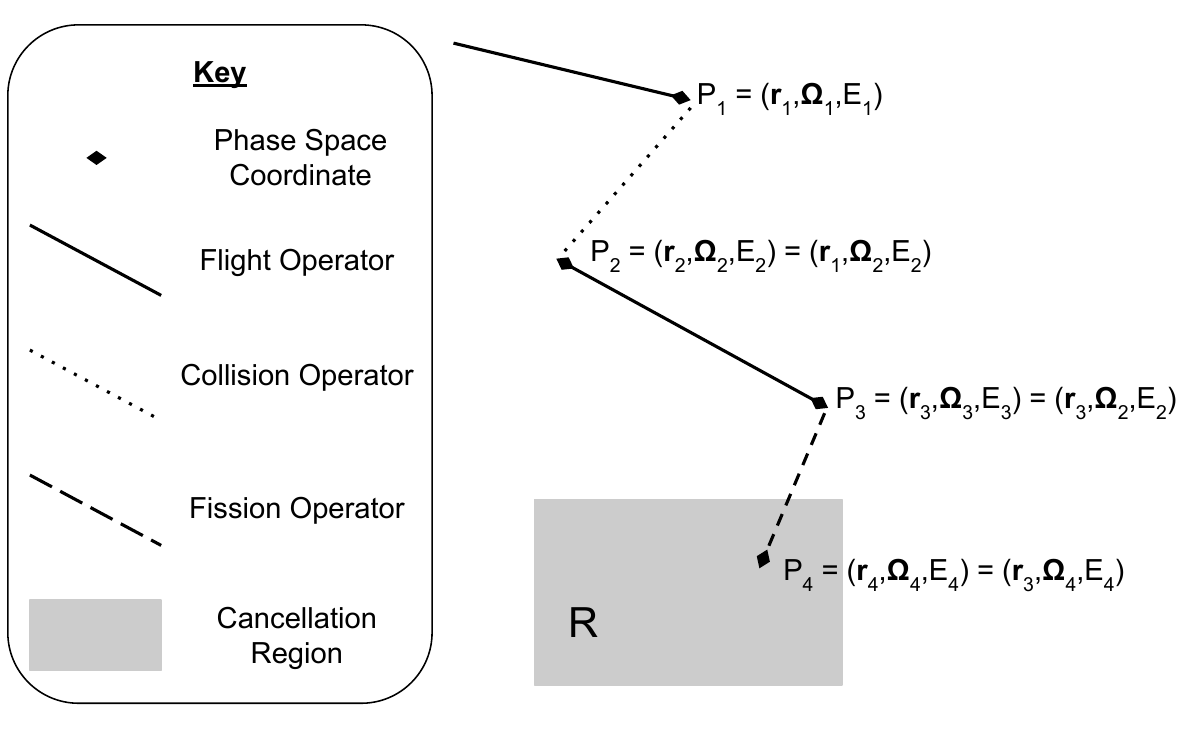}
            \end{minipage}\hfill%
            \begin{minipage}[c]{0.483\textwidth}
            \caption{Depicted here is the relationship between phase space points $P_1$, $P_2$, $P_3$, and $P_4$. Points connected by a flight operator (solid line) share the same direction and energy, and are only discontinuous in position. Points connected by a collision or fission operator (dotted and dashed lines respectively) share the same position, but are generally discontinuous in direction and energy. Any branches which might result from the application of the collision operator from $P_1$ to $P_2$ are not depicted.}
            \label{fig:collisions}
            \end{minipage}
        \end{figure*}

        Consider the following particle history. A neutron enters a collision at $P_1$, and then leaves that collision at $P_2$. The particle then undergoes a flight and experiences a fission at $P_3$. The fission at $P_3$ then contributes to the fission emission density at $P_4$. It is assumed that $P_4$ is located within the generalized phase space region $\mathcal{R}$, which will act as our cancellation region. \footnote{While our previous work in Ref.~\citenum{Belanger2021} used cancellations which only spanned space, we now consider cancellation regions spanning all dimensions of phase space. We therefore must consider three spatial dimensions, two dimensions for direction, and one dimension for energy.} This partial particle history is depicted in Fig.~\ref{fig:collisions}. Despite the fact that $\pos_4=\pos_3$, $P_3$ is not, in general, located in the cancellation region $\RR$, as $\dir_3$ and $E_3$ may not be within the domain of $\RR$. In order to examine the fission emission density at point $P_4$, we must first determine the collision density $\psi(P_3)$, for a given collision at $P_1$. From Eq.~\eqref{eq:psi} and Eq.~\eqref{eq:chi}, it follows that
        \begin{widetext} 
        \begin{equation}
            \psi(P_3) = \mathbb{T}\mathbb{C}\psi(P_3) = 
            \int T(P_2\to P_3) \int C(P_1\to P_2) \psi(P_1) \dd P_1 \dd P_2
        \end{equation}
        \begin{multline}
            = \iiint
            \dd\pos_1\dd\dir_1\dd E_1
            \iiint
            \dd\pos_2\dd\dir_2\dd E_2
            \psi\left(\pos_1,\dir_1,E_1\right)
            \nubar(\pos_1,E_1)
            \fbar\left(\dir_2,E_2 | \pos_1, \dir_1, E_1\right)
            \delta\left(\pos_2 - \pos_1\right) \\
            \Et\left(\pos_3,E_3\right)
            \exp\left(
                    -\int\limits_0^{\abs{\pos_3-\pos_2}} \Et\left(\pos_2 + u\dir_2,E_2\right)\dd u
                \right)
            \frac{\delta\left(\dir_3 - \dirto{\pos_2}{\pos_3}\right)
                  \delta\left(\dir_3-\dir_2\right)
                  \delta(E_3 - E_2)
                 }
                 {\abs{\pos_3 - \pos_2}^2}
        \end{multline}
        \begin{multline}
            = \iiint
            \dd\pos_1\dd\dir_1\dd E_1
            \psi\left(\pos_1,\dir_1,E_1\right)
            \nubar(\pos_1,E_1) \\
            \delta\left(\dir_3 - \dirto{\pos_1}{\pos_3}\right)
            \frac{\fbar\left(\dir_3,E_3|\pos_1,\dir_1,E_1\right)
                  \Et(\pos_3,E_3)
                 }
                 {\abs{\pos_3-\pos_1}^2}
            \exp\left(
                -\int\limits_0^{\abs{\pos_3-\pos_1}}
                \Et\left(\pos_1 + u\dir_3,E_3\right)\dd u
                \right)
                \text.
            \label{eq:psi_P3}
        \end{multline}
        The integral over $P_1$ in Eq.~\eqref{eq:psi_P3} indicates that $\psi(P_3)$ is a sum of
        contributions from all possible initial phase space points $P_1$ for which $\dir_3 =
        \dirto{\pos_1}{\pos_3}$.
        
        The fission emission density $\chi_f(P_4)$ is defined as
        \begin{equation}
            \chi_f(P_4) = \frac{1}{k}\mathbb{C}_f\psi(P_4)
            \text.
        \end{equation}
        Combining Eq.~\eqref{eq:col_fiss} and Eq.~\eqref{eq:psi_P3}, we obtain
        \begin{multline}
            \chi_f(P_4) =
            \iiint
            \dd\pos_1\dd\dir_1\dd E_1
            \psi\left(\pos_1,\dir_1,E_1\right)
            \nubar(\pos_1,E_1) \\
            \int
            \dd E_3
            \frac{\nu_f(\pos_4,E_3)
                  \Ef(\pos_4,E_3)
                  \ff\left(\dir_4,E_4|\pos_4,\dirto{\pos_1}{\pos_4},E_3\right)
                  \fbar\left(\dirto{\pos_1}{\pos_4},E_3|\pos_1,\dir_1,E_1\right) 
                 }
                 {k\abs{\pos_4-\pos_1}^2} \\
                 \exp\left(
                        -\int\limits_0^{\abs{\pos_4-\pos_1}}
                        \Et\left(\pos_1 + u\dirto{\pos_1}{\pos_4},E_3\right)
                        \dd u
                       \right) \\
            = 
            \iiint
            \psi\left(\pos_1,\dir_1,E_1\right)
            \nubar(\pos_1,E_1)
            \zeta(P_1\to P_4)
            \dd\pos_1\dd\dir_1\dd E_1
            \text.
            \label{eq:chi_P4}
        \end{multline}
        In the last step we have introduced the function $\zeta$:
        \begin{multline}
        \zeta(P_1\to P_4)= \int
            \dd E_3
            \frac{\nu_f(\pos_4,E_3)
                  \Ef(\pos_4,E_3)
                  \ff\left(\dir_4,E_4|\pos_4,\dirto{\pos_1}{\pos_4},E_3\right)
                  \fbar\left(\dirto{\pos_1}{\pos_4},E_3|\pos_1,\dir_1,E_1\right) 
                 }
                 {k\abs{\pos_4-\pos_1}^2} \\
                 \exp\left(
                        -\int\limits_0^{\abs{\pos_4-\pos_1}}
                        \Et\left(\pos_1 + u\dirto{\pos_1}{\pos_4},E_3\right)
                        \dd u
                       \right)
        \label{eq:fission_density_p1}
        \end{multline}
        \end{widetext}
        Here $\zeta(P_1\to P_4)$ is the transition kernel for a particle starting at $P_1$, undergoing a collision, then a flight, and then producing fission particles at $P_4$.
        
        We now wish to construct an estimator for the expected fission emission density at a point $Q\in\RR$. Our estimator operates on events where a fission particle is emitted at $P_4\in\RR$, from a particle originally entering a collision at $P_1$. To be unbiased, our estimator $\vartheta(P_1\to P_4|\RR,Q)$ for the fission emission density must have the property \cite{the_sacred_text}
        \begin{equation}
            \int \zeta(P_1\to P_4) \vartheta(P_1\to P_4|\RR,Q) \dd P_4 =
            \zeta(P_1\to Q)
            \text.
            \label{eq:zeta_theta}
        \end{equation}
        
        In order to achieve regional cancellation, we would like to define an estimator $\vartheta$ for the fission emission density at $Q$ where a portion of the fission emission density is located exactly at $Q$, and the remaining portion is uniformly distributed within the phase space region $\mathcal{R}$. We shall define this estimator to have the form
        \begin{equation}
            \vartheta_\eta(P_1\to P_4|\RR,Q) =
                (1-\eta)
                \delta(Q-P_4)
                +
                \frac{\eta}
                     {\mathcal{V}_\RR}
            \text.
            \label{eq:vartheta}
        \end{equation}
        Here, $\mathcal{V}_\RR$ is the generalized phase space volume occupied by $\RR$, and $\eta$ is the portion of the fission emission density that we wish to uniformly distribute within $\RR$.\footnote{Note that the parameter $\eta$ may take any value (real or complex); in particular, it is not required to lie in the $[0,1]$ interval.} If $\eta$ is taken to be a constant with respect to $P_4$, then, upon evaluation of the left-hand side of Eq.~\eqref{eq:zeta_theta}, using Eq.~\eqref{eq:vartheta}, we obtain:
        \begin{multline}
            \int \zeta(P_1\to P_4) \vartheta_\eta(P_1\to P_4 | \RR, Q) \dd P_4 = \\
            \int
            \zeta(P_1\to P_4)
            \left[
                (1-\eta)
                \delta(Q-P_4)
                +
                \frac{\eta}
                     {\mathcal{V}_\RR}
            \right]
            \dd P_4 =\\
            (1-\eta)\zeta(P_1\to Q) + \frac{\eta}{\mathcal{V}_\RR}\int\zeta(P_1\to P_4)\dd P_4
            \text.
            \label{eq:zeta_integral}
        \end{multline}
        Comparing Eq.~\eqref{eq:zeta_integral} and Eq.~\eqref{eq:zeta_theta}, it is clear that the only unbiased option is $\eta = 0$, corresponding to no cancellation. It is permissible however to allow $\eta = \eta(P_1,P_4)$ to be both a function of $P_1$ and $P_4$, as $\vartheta$ is already a function of these parameters. Using the ansatz
        \begin{equation}
        \eta(P_1,P_4) = \frac{\beta}{\zeta(P_1\to P_4)}
        \text,
        \label{eq:ansatz_eta}
        \end{equation}
        inspired by Eq.~\eqref{eq:w_u}, we see that
        \begin{widetext}
        \begin{multline}
            \int \zeta(P_1\to P_4) \vartheta_{\eta(P_1,P_4)}(P_1\to P_4 | \RR, Q) \dd P_4 = \\
            \int
            \zeta(P_1\to P_4)
            \left[
                \left(1-\frac{\beta}{\zeta(P_1\to P_4)}\right)
                \delta(Q-P_4)
                +
                \frac{\beta}
                     {\mathcal{V}_\RR\zeta(P_1\to P_4)}
            \right]
            \dd P_4 =\\
            \zeta(P_1\to Q) - \beta + \beta = \zeta(P_1\to Q)
            \text,
            \label{eq:zeta_theta_works}
        \end{multline}
        \end{widetext}
        which, compared with right-hand side of Eq.~\eqref{eq:zeta_theta}, shows that this choice leads to an unbiased estimator.

        Equation~\eqref{eq:zeta_theta_works} indicates that we are allowed to distribute a factor $\beta/\zeta(P_1\to P_4)$ of the fission particle uniformly within $\RR$, so long as $\beta$ has no functional dependence on $P_4$. This requirement on $\beta$ is essential to ensure that, after integrating over $P_4$, both $\beta$ terms will cancel; note however that $\beta$ is allowed to depend on $P_1$. We are therefore allowed to pick $\beta=0$ whenever it is convenient, so long as information from $P_4$ is not used to make this choice.
        
        Additionally, Eq.~\eqref{eq:zeta_theta_works} indicates that we must require $\zeta(P_1\to P_4) \neq 0\text{ }\forall P_4 \in \RR$. If this is not the case, then $\eta$ is undefined. In particular, this implies that we require $\Ef(\pos_4,E_3) > 0$ everywhere within our cancellation region. We must also require $\ff(\dir_4,E_4|\pos_4,\dirto{\pos_1}{\pos_4},E_3) > 0\text{ }\forall P_4\in\RR$; as fission is nearly perfectly isotropic, the angular component is not problematic, but the energy component could indeed be zero for very low energies, and care must therefore be taken when selecting the energy bounds for $\RR$. Despite these restrictions, we are given some liberty as to the definition of $\RR$, as it is allowed to be non-convex.
        
        The ansatz of Eq.~\eqref{eq:ansatz_eta} has the following remarkable property. Suppose that $\zeta(P_1\to P_4)$ has the structure
        \begin{equation}
            \zeta(P_1\to P_4)= h_0(P_1)\zeta_0(P_1\to P_4)
            \text.
        \end{equation}
        Consider the estimators
        \begin{align}
            \vartheta_\zeta &= \vartheta\left(P_1\to P_4|\RR,Q,\eta=\frac{\beta}{\zeta}\right)\\
            \vartheta_{\zeta_0} &= \vartheta\left(P_1\to P_4|\RR,Q,\eta=\frac{\beta_0}{\zeta_0}\right)
            \text.
        \end{align}
        The two estimators are actually identical for $\beta_0=\beta/h_0$. In other words, any factor in $\zeta(P_1\to P_4)$ that is independent of $P_4$ can be pulled out of the definition of $\zeta$ and still yield an unbiased estimator for the fission emission density.
        
        Finally, note that the integral over $E_3$ in the definition of $\zeta$ could be somewhat problematic and/or expensive to compute in a continuous-energy Monte Carlo code. It effectively corresponds to averaging over all possible nuclides, reaction channels, and energies, which could have been sampled in determining $P_3$, and leading to a fission particle at $P_4$. We therefore would like to determine if there is an alternative, simpler, unbiased option.
        
        \subsection{Expected-Value Estimators for Cancellation}
        \label{sec:expected_value_estimators}
        
        As it has been developed, $\zeta(P_1\to P_4)$ can be interpreted as a type of expected-value estimator \cite{spanierMonteCarloPrinciples2008}, because it is the expected contribution to the fission emission density at $P_4$, for a particle entering a collision at $P_1$. For the purpose of carrying out weight cancellation, several kinds of expected-value estimators for the fission emission density at $P_4$ could potentially be used in place of the form given by Eq.~\eqref{eq:fission_density_p1}. All that is required of $\zeta(P_1\to P_4)$ is that it be non-zero for all points $P_4$ in $\RR$. This is required by Eq.~\eqref{eq:zeta_theta_works}, as we can only distribute fission emission density uniformly within $\RR$ if $\zeta(P_1\to P_4)>0\text{ }\forall P_4\in\RR$. With this in mind, we will now consider what types of expected-value estimators could be used in lieu of $\zeta$.
        
        Next-event estimators are particular forms of expected-value estimators that average the sampled quantity over the following event in the stochastic process. Let us evaluate if a next-event estimator is suitable for the purpose of cancellation. Consider a next-fission estimator for the fission emission density; such an estimator is applied to particles undergoing a collision at $P_3=(\pos_3,\dir_3,E_3)$ and yields the expected fission emission density at a generic point $Q=(\pos,\dir,E)\in\RR$. Since fission does not change the position of particles, the contribution of the next-fission estimator vanishes everywhere except for $\pos=\pos_3$. Therefore, a next-fission estimator is not able to yield a non-vanishing contribution at all the points in cancellation region $\RR$.

        Thus, in order for cancellation to be possible, we need to include more than one event in our expected-value estimator, i.e.\ we need to use at least a next-next-event estimator, or possibly an estimator of even higher order. It is then crucial to determine the number of events that our estimator needs to look ahead and average over, in order to yield a non-vanishing contribution to all the phase space points in the cancellation region. Indeed, we want to \emph{minimize} the number of look-ahead events, because the evaluation of expected-value estimators becomes more and more cumbersome as the number of look-ahead events increases.
        
        Consider now the possibility of a next-flight-fission estimator. In our notation, such an estimator acts on particles emitted at $P_2$ and yields the expected fission emission density at $Q$, averaged over all possible flights from $P_2$ and all possible fission events. Since the flight operator does not modify the particle direction (see Eq.~\eqref{eq:flight_kernel}), the expected fission density vanishes everywhere except at positions reachable from $\pos_2$ with direction $\dir_2$. In general, this does not cover the whole cancellation region, except in the one-dimensional case \cite{Booth2010,Belanger2021}.
        
        It is now probably clear that a next-collision-flight-fission estimator should \emph{in general} yield a non-vanishing contribution to the fission emission density at all phase space points within $\RR$. In other words, given a particle undergoing a collision at $P_1$, the expected fission emission density (averaged over the next collision, flight, and fission) should not vanish anywhere within $\RR$. This corresponds to Eq.~\eqref{eq:fission_density_p1} above and justifies the construction of the previous section.
        
        Two remarks are in order here. First, there are cases where even a next collision-flight-fission estimator is not sufficient to achieve a non-vanishing expected fission emission density at all points within $\RR$. Indeed, the collision between $P_1$ and $P_2$ may be subject to kinematic constraints, and Eq.~\eqref{eq:fission_density_p1} shows that the expected fission emission density vanishes if $\fbar\left(\dirto{\pos_1}{\pos},E_3|\pos_1,\dir_1,E_1\right)=0$ for some $\pos$ in cancellation region $\RR$. Second, an estimator based on Eq.~\eqref{eq:fission_density_p1} would require the evaluation of the integral over $E_3$ at every collision, which is impractical. In fact, regional cancellation attempts to perform the cancellation algorithm a posteriori, after flights have already been sampled, and fission particles have been produced. Since we have already sampled a Monte Carlo history from $P_1$ through $P_2$, $P_3$, and $P_4$, we would like to reuse as much information as possible from the sampled history to remove part of the fission density from $P_4$ and redistribute it uniformly within $\RR$.
        
        Thus, our expected-value estimator needs to average over sufficiently many event samplings to be able to ``see'' the whole region $\RR$; at the same time, we want our estimator to average over the strict minimum number of samplings. Each additional real variable that we average over introduces an extra integration in the expression of the expected fission emission density and reduces the usefulness of the $P_1\to P_4$ history that we have already sampled.

        \subsection{Intermediate Collision Points}
        \label{sec:condition_on_sampled_scatter}

        For the subsequent analysis, it is useful to consider a different form of the collision kernel, more aligned with how most continuous-energy Monte Carlo codes sample a collision event. While Eq.~\eqref{eq:C_kernel_simp} presents the collision kernel in terms of the averaged macroscopic cross sections and yields, most continuous-energy Monte Carlo codes do not handle collisions in such a manner. In production-level codes, microscopic cross sections are tabulated for different nuclides and different reaction channels (elastic, level inelastic, etc.) \cite{MCNP,T4,OpenMC,Serpent}. Each combination of nuclide and reaction channel has an independent transfer function for each type of non-capture collision. The concentration $N_i(\pos')$ of nuclide $i$ is a function of position, and the total microscopic cross section $\sigma_i(\pos',E')$ is a function of position and energy.\footnote{The microscopic cross section is typically given as a function of temperature and energy. However, since the temperature is a function of position, we have chosen to present the microscopic cross section as a function of position and energy, to avoid the introduction of a superfluous variable.} The total macroscopic cross section is
        \begin{equation}
            \Et(\pos',E') = \sum_i N_i(\pos')\sigma_i(\pos', E')
            \text.
        \end{equation}
        At a collision site, we select the nuclide with which our particle will undergo a collision: nuclide $i$ is chosen with probability $N_i(\pos')\sigma_i(\pos',E')/\Et(\pos',E')$. With nuclide $i$ having been sampled, a reaction channel $m$ must next be sampled. If we let $\sigma_{i,m}(\pos',E')$ be the partial microscopic cross section for channel $m$, then the total microscopic cross section is
        \begin{equation}
            \sigma_i(\pos',E') = \sum_m \sigma_{i,m}(\pos',E')
            \text,
        \end{equation}
        and channel $m$ will be selected with probability $\sigma_{i,m}(\pos',E')/\sigma_{i}(\pos',E')$. This channel has an associated yield of $\nu_{i,m}(E')$, and transfer function $f_{i,m}\left(\dir,E|\dir',E'\right)$. Continuous-energy nuclear data files typically give $f_{i,m}$ as a product of a marginal PDF in energy and a conditional PDF in direction:
        \begin{multline}
            f_{i,m}\left(\dir,E|\dir',E'\right) = \\
            f_{i,m}\left(E|\dir',E'\right)
            f_{i,m}\left(\dir|\dir',E',E\right)
            \label{eq:transfer_marginal}
            \text.
        \end{multline}
        When this is the case, the energy $E$ is first sampled from the marginal PDF, and the direction is subsequently sampled from the conditional PDF. With these provisions, it is then possible to write the collision kernels, Eqs.~\eqref{eq:col_scat} and \eqref{eq:col_fiss}, as
        \begin{widetext}
        \begin{subequations}
        \begin{align}
            C_s(P'\to P) =
            \frac{\delta\left(\pos-\pos'\right)}{\Et(\pos',E')}
            \sum_i N_i(\pos')
            \sum_{\substack{m\\m\not = \text{fiss}}}
            \nu_{i,m}(E')
            \sigma_{i,m}(\pos',E')
            f_{i,m}\left(E|\dir',E'\right)
            f_{i,m}\left(\dir|\dir',E',E\right)\\
            C_f(P'\to P) =
            \frac{\delta\left(\pos-\pos'\right)}{\Et(\pos',E')}
            \sum_i N_i(\pos')
            \nu_{i,\text{fiss}}(E')
            \sigma_{i,\text{fiss}}(\pos',E')
            f_{i,\text{fiss}}\left(E|\dir',E'\right)
            f_{i,\text{fiss}}\left(\dir|\dir',E',E\right)
            \text.
        \end{align}
            \label{eq:C_kernel_CE}
        \end{subequations}
        \end{widetext}

        Based on the form of Eqs.~\eqref{eq:C_kernel_CE}, we introduce then the concept of an ``intermediate collision point'', indicating that the required pieces of information are sampled incrementally when performing a collision. Examples of intermediate collision points would be the state where we have sampled only the nuclide, or the nuclide and the channel, or the nuclide, reaction channel, and energy. With the concept of an intermediate collision, it is then evident that there is an intermediate collision point between $P_1$ and $P_2$, where the particle has selected an isotope to collide with ($i$), a reaction channel ($m$), and even an outgoing energy ($E_2$), but has yet to select a direction $\dir_2$ out of the collision. In general, this intermediate collision point between $P_1$ and $P_2$ is the strict minimum number of steps we must look back in a particle's history, in order to see a nonzero fission emission density everywhere within $\RR$ (assuming that is is possible to scatter into all directions subtended by $\RR$). This state is accessible in a Monte Carlo simulation, as the nuclide, reaction channel, and energy $E_3$ were all sampled when producing the fission particle at $P_4$, and this information can be stored with the particle. The transition kernel from $P_1$ to $P_4$, given a collision with nuclide $i$ in reaction channel $m$ and outgoing energy $E_3$, is then
        \begin{widetext}
        \begin{multline}
            \zeta(P_1\to P_4| i, m, E_3) =
            \frac{\nu_f(\pos_4,E_3)\Ef(\pos_4,E_3)
                  \ff\left(\dir_4,E_4|\pos_4,\dirto{\pos_1}{\pos_4},E_3\right)
                  f_{i,m}\left(\dirto{\pos_1}{\pos_4}|\dir_1,E_1,E_3\right)
                 }
                 {k\abs{\pos_4 - \pos_1}^2}\times \\
            \exp\left(
                 -\displaystyle\int\limits_0^{\abs{\pos_4 - \pos_1}}
                 \Et\left(\pos_1 + u\dirto{\pos_1}{\pos_4},E_3\right)\dd u
                \right)
            \text.
            \label{eq:zeta_conditioned}
        \end{multline}
        \end{widetext}
        This is now quite reminiscent of the fission density function which we used in our previous work \cite{Belanger2021}, as summarized in Sec.~\ref{sec:summary_previous_paper} (see Eq.~\eqref{eq:previous_paper_zeta}). It is worth stressing that Eq.~\eqref{eq:zeta_conditioned} uses the macroscopic fission cross section $\Ef$ and the average fission transfer function $f_\text{fiss}$, which are averaged over all fissile nuclides at $\pos_4$. In general, these quantities might vary within $\RR$, due to spatial dependence in the nuclide concentrations and temperature. Examining under what circumstances the estimator $\vartheta(P_1\to P_4|\RR,Q,\eta)$ is unbiased for the transition kernel presented in Eq.~\eqref{eq:zeta_conditioned}, it is straightforward to observe that this condition is met for the choice of $\eta = \beta/\zeta\left(P_1\to P_4| i, m, E_3\right)$.
        
        Equation~\eqref{eq:zeta_conditioned} (and its associated estimator) are subject to the same constraints as Eq.~\eqref{eq:fission_density_p1}, as discussed in Sec.~\ref{sec:expected_value_estimators}: namely, the expected fission emission density must be non-zero at all the points in the cancellation region. In particular, it is required that $f_{i,m}\left(\dirto{\pos_1}{\pos_4}|\dir_1,E_1,E_3\right) > 0\text{ }\forall\text{ }\pos_4\in\RR$. A special case arises when the reaction channel $m$ uses a delta distribution for either the energy or direction (such as in level inelastic scattering). If such a channel was selected during the last collision, then the value of $\zeta(P_1\to P_4)$ is infinite at the collision point (as we are evaluating the delta distribution at the singularity), and it vanishes almost everywhere within $\RR$. Thus, such channels do not generally partake in cancellation, as the uniform portion would then necessarily be zero according to Eq.~\eqref{eq:ansatz_eta}. Finally, not all nuclear data facilitates the decomposition provided by Eq.~\eqref{eq:transfer_marginal}. Sometimes the joint PDF might be provided as a marginal PDF in direction, and a conditional PDF in energy. If this is the case, one must go back to the intermediate collision point before having sampled the direction, in order to see the entire cancellation region. 

        If the region $\RR$ contains only one material, which is completely homogeneous in nuclide concentrations and temperature, then $\nu_f$ and $\Ef$ are independent of $P_4$. As we have discussed in Sec.~\ref{sec:estimators_averaging_over_all}, these factors may be removed from the definition of $\zeta$, without compromising the unbiasedness of the method. In addition, if fission is assumed to be perfectly isotropic (a frequent assumption), and if the fission energy $E_4$ is completely independent of the incident energy and direction, then we do not actually need to perform cancellation on the fission emission density, but only on the collision density at $\pos_4$. For the multi-group benchmark in Ref.~\citenum{Belanger2021} in which we previously demonstrated exact regional cancellation in 3D, we used homogeneous cancellation regions, where both fission and scattering were isotropic, and the fission energy was also assumed to be independent of the incident energy $E_3$. This indicates that $\zeta$ could be simplified to
        \begin{multline}
            \zeta(P_1 \to P_4| E_3) = \\
            \frac{\exp\left(-\displaystyle\int\limits_0^{\abs{\pos_4-\pos_1}}\Et\left(\pos_1 + u\dirto{\pos_1}{\pos_4},E_3\right)\dd u\right)}{\abs{\pos_4 - \pos_1}^2}
        \end{multline}
        and still result in an unbiased cancellation method.
        
        \subsection{Non-Uniform Cancellation}
        \label{sec:non-uniform}
        
        Suppose now that, instead of distributing some fission emission density uniformly over $\RR$, we would like to distribute fission emission density according to an arbitrary function, $D(Q)$. In this case, our estimator must be modified as 
        \begin{multline}
            \vartheta_D(P_1\to P_4|\RR,Q,\eta) = \\
                \frac{\zeta(P_1\to Q|i,m,E_3) - D(Q)\beta}
                     {\zeta(P_1\to Q|i,m,E_3)}
                \delta(Q-P_4)
                + \\
                \frac{\beta}
                     {\zeta(P_1\to P_4|i,m,E_3)}
                \frac{D(Q)}
                     {\mathcal{V}_\RR}
            \text.
            \label{eq:vartheta_non_unifomrm}
        \end{multline}
        This can be shown by following the same approach taken in Eq.~\eqref{eq:zeta_theta_works}.
        
        Here, we have placed a portion $(\zeta(P_1\to Q|i,m,E_3) - D(Q)\beta) / \zeta(P_1\to Q|i,m,E_3)$ of the weight at the sampled point $P_4$, and a portion $\beta/\zeta(P_1\to Q|i,m,E_3)$ of the particle is distributed according to $D(Q)$. It is only assumed here that $D(Q)$ is dimensionless and that it is piece-wise continuous. In theory, there is no reason that $D(Q)$ could not be negative, or even complex valued\footnote{While this paper only considers particles with a single real-valued statistical weight, some transport problems require that particles carry a complex weight \cite{Yamamoto2012, Rouchon2017}.}; if $D(Q)$ is negative or complex, it might not be possible to sample it directly, but such a situation might be treated using e.g.\ importance sampling \cite{the_sacred_text}. Finally, we note that for the choice of $D(Q) = 1$ the case of uniform cancellation is retrieved.
        
        \subsection{Fission Emission Density Function with Delta-Tracking Schemes}
        \label{sec:cancellation_and_dt}
        
        In the above derivations, we have often made use of the non-homogeneous exponential
        distribution
        \begin{equation}
            \Et(\pos+d\dir,E)\exp\left(-\int\limits_0^{d}\Et(\pos+s\dir,E)\dd s\right)
            \label{eq:exponential_integral}
            \text,
        \end{equation}
        which occurs in the flight kernel $T(P'\to P)$ and in the transition kernel $\zeta(P'\to P)$. This distribution is sampled when trying to determine the distance $d$ a particle will travel from initial position $\pos$ along direction $\dir$, before undergoing a collision. For the case of piece-wise constant macroscopic cross sections, this distribution is straightforward to sample for $d$. However, when the macroscopic cross section is not piece-wise constant, more sophisticated methods than direct sampling are often employed \cite{Belanger2020}. Delta-tracking and negative-weighted delta-tracking are two such methods that sample the distance to collision using a sampling cross section $\Sigma_\text{smp}(E)$, and then sample whether a real or virtual collision has occurred with a specific criterion \cite{Leppanen2017,Legrady2017,Carter1972}; in the distinct case of negative-weighted delta-tracking, a weight modifier may be additionally applied to the particle's weight, which could potentially be negative \cite{Legrady2017,Carter1972}. \footnote{Using negative weights can be advantageous in some cases, as it allows $\Sigma_\text{smp}(E)$ to be less than $\Et(\pos,E)$. In delta-tracking, it is required that $\Sigma_\text{smp}(E)\ge\Et(\pos,E)$ everywhere in the problem domain: because of this requirement, it could be difficult to determine $\Sigma_\text{smp}(E)$ for delta-tracking, when considering spatially continuous cross sections \cite{Belanger2020}.} In a real collision, the particle undergoes a reaction as normal. In a virtual collision, the particle's energy and direction do not change; this event is usually known as \emph{delta scattering}. The particle continues to sample new flight distances and to move to the new location, until a real collision is sampled. Coleman \cite{Coleman1968} and Legrady et al.\ \cite{Legrady2017} have previously provided evidence as to why such sampling methods are unbiased.
        
        It is possible to include virtual collisions in the transport equations by modifying the flight kernel to be
        \begin{multline}
            T_{DT}(P'\to P) = \\
            \frac{\Sigma_\text{smp}(E')
                  \exp\left(
                    -\Sigma_\text{smp}(E')\abs{\pos-\pos'}
                  \right)
                 }
                 {\abs{\pos-\pos'}^2} \\
            \delta\left(\dir - \dirto{\pos'}{\pos}\right)
            \delta\left(\dir'-\dir\right)
            \delta\left(E-E'\right)
            \text,
            \label{eq:DT_flight_kernel}
        \end{multline}
        and the collision kernel to be
        \begin{multline}
            C_{DT}(P'\to P) = \\
            \frac{\Et(\pos',E')}{\Sigma_\text{smp}(E')} C(P'\to P) + \\
            \left(1-\frac{\Et(\pos',E')}{\Sigma_\text{smp}(E')}\right)
            \delta(E-E')\delta(\dir - \dir')\delta(\pos-\pos')
            \text.
            \label{eq:DT_collision_kernel}
        \end{multline}
        These equations are valid for both delta tracking and negative-weighted delta tracking.
        From Eq.~\eqref{eq:DT_flight_kernel}, the PDF for leaving a collision site at $\pos'$ and flying to $\pos$ and inducing a fission (given that we are flying in the direction of $\pos$, i.e.\ $\dir'=\dirto{\pos'}{\pos}$) is
        \begin{equation}
            \Ef(\pos,E')\exp\left(-\Sigma_\text{smp}(E')\abs{\pos-\pos'}\right)
            \text,
        \end{equation}
        which is exactly the form presented in Eq.~\eqref{eq:previous_paper_zeta}. Thus, delta-tracking-like algorithms provide the large advantage of not requiring the integration of the total cross section along the flight path. This makes them interesting for the purpose of performing exact regional cancellation.

        The form of Eq.~\eqref{eq:DT_flight_kernel} is valid regardless of whether the collision at $\pos'$ was real or virtual. Equation~\eqref{eq:DT_collision_kernel} shows that the angular distribution for virtual collisions is singular, because it is described by a delta distribution. As discussed in Sec.~\ref{sec:condition_on_sampled_scatter}, channels with singular distributions are not allowed to partake in cancellation, i.e.\ we need to set $\beta=0$ for all such channels. For the particular case of virtual collisions, however, another treatment is possible. At the site where the virtual collision took place, there was a probability that the particle could have instead undergone a real collision. We can therefore imagine ``splitting'' the particle before the collision. A weight $w(1-\Et(\pos,E')/\Sigma_\text{smp}(E'))$ is considered to undergo a virtual collision, and have its next collision at $P_3$; this virtual collision portion cannot be used in cancellation, as the angular distribution was a delta distribution, and the uniform component is then always zero, as explained in Sec.~\ref{sec:condition_on_sampled_scatter}. The rest of the particle weight, namely $w\Et(\pos,E')/\Sigma_\text{smp}(E')$, is considered to undergo a real collision and have its next collision at $P_3$, like the virtual part. However, this part can also partake in cancellation. The point-wise fission particle weight which must remain at the sampled fission particle site is then
        \begin{align}
            &w\left(1-\frac{\Et(\pos,E')}{\Sigma_\text{smp}(E')}\right) + w\frac{\Et(\pos,E')}{\Sigma_\text{smp}(E')}\left(1-\frac{\beta}{\zeta(P'\to P)}\right) = \nonumber \\
            &w\left(1-\frac{\beta}{\zeta(P'\to P)}\frac{\Et(\pos,E')}{\Sigma_\text{smp}(E')}\right) = \nonumber \\
            &w\left(1-\frac{\beta'}{\zeta(P'\to P)}\right) 
            \text,
        \end{align}
        where we have set $\beta'=\beta\Et/\Sigma_\text{smp}$. Thus, splitting shows that \emph{all} collisions can be assumed to partake in cancellation as if they were real, because the presence of virtual collisions only affects the choice of $\beta$. Since the estimator is unbiased for any $\beta$, the factor $\Et(\pos,E')/\Sigma_\text{smp}(E')$ is not necessary. However, note that this approach is only unbiased so long as at the virtual collision site the real component of the scattering kernel for forward scattering with no energy change is not zero (i.e.\ $C(P\to P)\not = 0$). Otherwise, the real collision component could not reach $P_3$, as it would be impossible to have a real collision with forward scattering and no change in energy. This was possible in our previous multi-group example, because in-group scattering was always allowed and all scattering was assumed to be isotropic; however, this might not be as trivial in a continuous-energy setting.

    \section{Optimization of Cancellation Efficiency}
    \label{sec:optimization}
         
        We now turn our attention to the optimal choice of the free parameter $\beta$ of the cancellation estimator, used to calculate the weight that can be uniformly distributed over the cancellation region. In Booth and Gubernatis's seminal paper \cite{Booth2010} and in our previous work \cite{Belanger2021}, $\beta$ was chosen to be the minimum of the expected fission density over the cancellation region. This choice has the advantage of being relatively easy to evaluate, but it is not necessarily the most efficient one. In this section we attempt to introduce a better strategy to determine the cancellation parameter $\beta$ for each particle partaking in cancellation.

        In order to optimize for the cancellation efficiency, one must first properly define the quantity to be optimized. As we mentioned in Sec.~\ref{sec:summary_previous_paper}, the maximum amount of cancellation will occur when the sum of the absolute value of all the weights remaining after cancellation in the region has been minimized. For $N$ particles which initially land in a cancellation region, we define the absolute value of all weight in a region after cancellation as
        \begin{equation}
            \Gamma_1 = \sum\limits_{k=1}^{N}\abs{w_{k,p}} +
            \abs{\sum\limits_{k=1}^{N}w_{k,u}},
            \label{eq:gamma1}
        \end{equation}
        with $w_{k,p}$ being the point-wise weight of particle $k$, and $w_{k,u}$ the uniform weight portion of particle $k$.\footnote{While there are $N$ particles in the cancellation region before the cancellation operations have been carried out, there will be more than $N$ particles after cancellation, due to the new uniform particles which are created during the cancellation process.} Equation~\eqref{eq:gamma1} is the total post-cancellation weight discussed in Sec.~\ref{sec:summary_previous_paper}, Eq.~\eqref{eq:post_cancellation_total_weight}. As each particle has a different value for the cancellation parameter $\beta$, we then may substitute to obtain
        \begin{equation}
            \Gamma_1 = \sum\limits_{k=1}^{N}\abs{\frac{\zeta_k - \beta_k}{\zeta_k}w_{k}} +
            \abs{\sum\limits_{k=1}^{N}\frac{\beta_k}{\zeta_k}w_{k}}
            \text,
        \end{equation}
        where $w_k$ and $\beta_k$ are respectively the pre-cancellation weight and the cancellation parameter of the $k$-th particle, and $\zeta_k = \zeta(P'_k \rightarrow P_k)$
        is the expected fission density of the $k$-th particle, which is assumed to have had its previous collision in $P'_k$ and its fission event in $P_k$ (note that we have simplified the notation here compared to Sec.~\ref{sec:unbiasedness}; for a given particle, $P'_k$ and $P_k$ respectively correspond to $P_1$ and $P_4$).

        Due to the presence of the
        absolute values, it is quite difficult to optimize the expression of $\Gamma_1$ analytically with respect to $\beta_k$. We instead
        define a modified quantity $\Gamma_2$, which shares a minimum with $\Gamma_1$:
        \begin{equation}
            \Gamma_2 = \sum\limits_{k=1}^{N}\bigg(\frac{\zeta_k - \beta_k}{\zeta_k}w_{k}\bigg)^2
            + \bigg(\sum\limits_{k=1}^{N}\frac{\beta_k}{\zeta_k}w_{k}\bigg)^2.
            \label{eq:L2_norm}
        \end{equation}
        We now wish to obtain the set of optimal values $\beta_k$ that minimize $\Gamma_2$. To remain unbiased, we are not allowed to calculate $\beta_k$ based on the phase space coordinates $P_k$ where the particle $k$ landed in the cancellation region (this was made evident in Eq.~\eqref{eq:zeta_theta_works}). As $\zeta_k = \zeta(P'_k \rightarrow P_k)$, we cannot directly minimize Eq.~\eqref{eq:L2_norm}. In the two subsequent sections, we will present two reasonable options to avoid this problem, both possibly giving way to a method of optimizing the regional cancellation algorithm. We go through the optimization for each case, obtaining two different formulations for calculating the set of optimal values for $\beta_k$.
        
       \subsection{Replacing $\zeta_k$ with $\expval{\zeta_k}$}\label{optimization_replacement}
       
            The first approach consists in averaging $\zeta_k$ over the entire phase space of the region $\mathcal{R}$, such that
            \begin{equation}
                \expval{\zeta_k} = \frac{\displaystyle\int_\mathcal{R}\zeta(P'_k \rightarrow P_k)\dd P_k}{ \displaystyle\int_\mathcal{R}\dd P_k}.
                \label{eq:expval_f}
            \end{equation}
            We may then replace $\zeta_k$ with $\expval{\zeta_k}$ in Eq.~\eqref{eq:L2_norm}, and optimize the new approximate form
            \begin{equation}
                \Gamma^*_{2} = \sum_{k=1}^{N}\bigg(\frac{\expval{\zeta_k} - \beta_k}{\expval{\zeta_k}}w_k\bigg)^2 + \bigg(\sum_{k=1}^{N}\frac{\beta_k}{\expval{\zeta_k}}w_k\bigg)^2
                \text,
                \label{eq:L2_mod}
            \end{equation} 
            which is now independent of $P_k$. The detailed derivation for this approach is presented in Appendix~\ref{ap:opt_derivation}, and the resulting equation for the cancellation parameter $\beta_k$ is found to be
           \begin{equation}
               \beta_k = \expval{\zeta_k}\left(1- \frac{S^*}{w_k}\right),
                \label{eq:opt_beta_j}
           \end{equation}
           where we make use of the definition
           \begin{equation}
               S^* = \frac{W}{N+1},
               \label{eq:S_tilde}
           \end{equation}
           and
           \begin{equation}
               W = \sum_{k=1}^{N} w_k,
               \label{eq:W}
           \end{equation}
           $W$ being the net weight in the region $\mathcal{R}$ before cancellation. It should also be mentioned that Eq.~\eqref{eq:opt_beta_j} would also be obtained if we first minimized Eq.~\eqref{eq:L2_norm} with respect to $\beta_k$ and then averaged over $P_k$.
           
        \subsection{Optimization of $\expval{\Gamma_2}$}\label{optimization_average}
        
            The second approach consists in averaging $\Gamma_2$ over the phase space of the region $\mathcal{R}$, obtaining
            \begin{equation}
                \expval{\Gamma_2} = \frac{\displaystyle\int_\mathcal{R}\Gamma_2\prod\limits_{k=1}^{N}\zeta_k\dd P_k} {\displaystyle\int_\mathcal{R}\prod\limits_{k=1}^{N}\zeta_k\dd P_k}.
                \label{eq:avg_l2}
            \end{equation}
            We may then optimize $\expval{\Gamma_2}$ instead of $\Gamma_2$. The complete derivation is provided in Appendix~\ref{ap:opt2_derivation}, and leads to a different equation for $\beta_k$:
            \begin{equation}
                \beta_k = \expval{\zeta_k}c_k\bigg(1-\frac{S}{w_k}\bigg).
                \label{eq:opt2_beta_j}
            \end{equation}
            Here, we have made use of the two following definitions:
            \begin{equation}
               c_k = \bigg(2\expval{\zeta_k}\expval{\frac{1}{\zeta_k}}-1\bigg)^{-1},
               \label{eq:ck}
            \end{equation}
            where the angle brackets have the same meaning as in Eq.~\eqref{eq:expval_f}, and
            \begin{equation}
                S = \frac{\displaystyle\sum_{k=1}^N c_k w_k}{1+\displaystyle\sum_{k=1}^N c_k}.
                \label{eq:S}
            \end{equation}
            
       \subsection{Small Region Limit}
       
           For the two possible methods that we have outlined to minimize the weight after cancellation, we are left with two different possibilities for the value of $\beta_k$. At first glance, these two choices of $\beta_k$ look quite different. Upon closer inspection of the definition of $c_k$ in Eq.~\eqref{eq:ck}, we notice that $c_k=1$ only if $1/\expval{\zeta_k}=\expval{1/\zeta_k}$. When this is the case, it then follows from Eq.~\eqref{eq:S} that $S=S^*$. This then indicates that the two definitions of $\beta_k$ given by Eq.~\eqref{eq:opt_beta_j} and Eq.~\eqref{eq:opt2_beta_j} are equivalent only when $1/\expval{\zeta_k}=\expval{1/\zeta_k}$. In general, however, Jensen's inequality implies $1/\expval{\zeta_k}\leq\expval{1/\zeta_k}$ \cite{Feller}. If a cancellation region were defined such that it is small enough that one could reasonably assume that $\zeta(P'_k\rightarrow P_k)$ is nearly constant within the region, then $c_k\approx 1$, leading to the two methods being equivalent. However, this would likely require such a small region that it is very unlikely that any other particles would be located within the region, which will make cancellation very ineffective.
           
    \section{Monte Carlo Implementation}\label{sec:implementation}    
        \subsection{Cancellation with Distributed Memory Simulations}
        
            Most production Monte Carlo codes make use of distributed-memory parallel computing techniques such as Message Passing Interface (MPI), although the exact algorithm used varies from code to code \cite{T4,MCNP,OpenMC,Serpent}. Generally speaking, distributed-memory parallelization can pose a problem for cancellation, which is by construction more efficient when there are more particles in each cancellation region. With distributed-memory parallelization, the fission particles within a given cancellation region will be distributed amongst several nodes. To ensure the highest possible efficiency, cancellation must be performed on the entire fission source. One method to do this is to send all of the fission particles to the master node between power iteration generations, and then perform cancellation only on the master node. Another option would be to use a method inspired by domain decomposition \cite{Brunner2009}, where certain nodes perform cancellation for certain regions, and fission particles would need to be sent to the node which corresponds to their cancellation region. This method is certainly possible, but likely much more difficult to implement in production Monte Carlo codes. For the proof of concept presented in this paper, we have chosen to use the former method, sending all fission particles to the master node for cancellation.
    
        \subsection{Calculation of $\expval{\zeta_k}$ and $\expval{1/\zeta_k}$}
        
            In order to evaluate $\beta_k$ according to Eq.~\eqref{eq:opt_beta_j}, we must have knowledge of $\expval{\zeta_k}$. For Eq.~\eqref{eq:opt2_beta_j}, we additionally need knowledge of $\expval{1/\zeta_k}$. In general, it is not possible to analytically calculate either of these quantities for particle $k$, born with phase space coordinates $P_k$ located within cancellation region $\mathcal{R}$. However, it is possible to estimate both of these quantities with a Monte Carlo sampling approach. For each particle $k$, we know the phase space coordinates $P'_k$ of its previous collision, and we know the bounds of the outgoing phase space coordinates $P_k$ which define the cancellation region $\mathcal{R}$.
            
            Assume that a set of non-overlapping, hypercuboid cancellation regions are imposed on top of the problem domain. Then, between each generation of power iteration, the fission particles (having stored their parent's previous phase space coordinates $P'_k$) may be sorted into the cancellation regions, based on their phase space coordinates $P_k$. Once this is accomplished for a given cancellation region $\mathcal{R}$, we may iterate over all particles in $\mathcal{R}$, and estimate their values of $\expval{\zeta_k}$ and additionally $\expval{1/\zeta_k}$, depending on which optimisation algorithm is chosen. The estimates for these quantities may be obtained using
            \begin{equation}
                \expval{\zeta_k} \approx \frac{1}{N_s}\sum_{i=1}^{N_s} \zeta(P'_k \rightarrow \tilde{P}_i)
                \label{eq:exp_fk}
            \end{equation}
            and
            \begin{equation}
                \expval{\frac{1}{\zeta_k}} \approx \frac{1}{N_s}\sum_{i=1}^{N_s} \frac{1}{\zeta(P'_k \rightarrow \tilde{P}_i)},
                \label{eq:exp_1_fk}
            \end{equation}
            respectively, where $N_s$ is the number of samples to be used in the estimation, and the outgoing phase space coordinates $\tilde{P}_i$ are pseudo-randomly sampled so that $\tilde{P}_i \in \mathcal{R}$ and $\tilde{P}_i \sim \mathcal{U}(\mathcal{R})$. This is straightforward to accomplish with cuboid regions.
            
            With this approach, a better estimate of $\expval{\zeta_k}$ and $\expval{1/\zeta_k}$ may be obtained by augmenting the number of samples. As $N_s$ is increased, the error on the estimate of the two expectation values will decrease according to $\mathcal{O}(1/\sqrt{N_s})$ \cite{Niederreiter1978}. This indicates that a large $N_s$ may be required to obtain a suitable estimate of $\expval{\zeta_k}$ and $\expval{1/\zeta_k}$. Even more problematic is the fact that evaluating $\zeta(P'_k\rightarrow \tilde{P}_i)$ could be quite costly; this is especially true in the case of continuous-energy neutron transport problems, where many evaluations of scattering distributions would be necessary. In order to reduce $N_s$, while still obtaining adequate estimates for $\expval{\zeta_k}$ and $\expval{1/\zeta_k}$, we propose the use of a quasi-random technique, using a Sobol' sequence \cite{Niederreiter1978} to sample the outgoing phase space coordinates $\tilde{P}_i$. This approach generally has a better convergence rate than using a pseudo-random number generator to sample $\tilde{P}_i$, as it leads to a more uniform exploration of the phase space \cite{Niederreiter1978}. At any rate, we stress that statistical uncertainties on the estimation of $\expval{\zeta_k}$ and $\expval{1/\zeta_k}$ only affect the efficiency of the cancellation method, as we have proved that the method is unbiased for any values of the free parameters $\beta_k$.
            
        \subsection{Heterogeneous Cancellation Regions}\label{heterogeneous_regions}
        
            In this work, we have proposed two possible  approaches to selecting an optimal value of $\beta$, to optimize the amount of weight which is cancelled. Neither of these approaches requires the minimum value of the fission emission density within the cancellation region $\RR$. Hence, it is no longer necessary to restrict the cancellation regions to be cuboids, as we were required to do in Ref.~\citenum{Belanger2021}. In light of our proposed sampling methods to estimate $\expval{\zeta_k}$ and $\expval{1/\zeta_k}$ to obtain the optimized cancellation parameter $\beta_k$, it is evident that a rejection technique may be applied to isolate different material regions within a given cuboid cancellation region. If we have a cancellation region with fuel and water, then all of the fission particles are of course only born inside the fuel portion, and the fission density everywhere in the water should be zero. When sampling the random phase space coordinates $\tilde{P}_i$, we must now add the requirement that $\tilde{P}_i$ have spatial coordinates that are located inside of the fuel material.
            
            Other special cases can be handled using this approach. For example, cancellation can also be performed when there are two non-connected fuel regions within the same cancellation mesh region. We may also have cancellation regions which contain multiple different fuel regions. Using rejection sampling to determine cancellation regions makes it easy to apply regional cancellation to complex geometries encountered in realistic reactor physics problems. Of course, the rejection procedure must also be applied to the sampling of the phase space coordinates of the uniformly distributed particles.
            
        \subsection{Monte Carlo Implementation in the open-source code MGMC}
            
            For our previous work on regional cancellation, a multi-group Monte Carlo mini-app called MGMC was used to test cancellation on a well-known reactor physics benchmark. MGMC has been developed to facilitate the fast and easy implementation and testing of new transport algorithms. Being only $\approx 13\text{ k}$ lines of code, it is much faster to test new ideas in MGMC than it would be in a large industrial code. General 3D geometries are supported using a standard constructive solid geometry formalism based on surfaces, universes, and lattices, familiar to any user of other well-known Monte Carlo codes \cite{T4,MCNP,Serpent,OpenMC}. Different mesh tallies are available for flux or reaction rates, with track-length or collision estimators. MGMC can solve fixed-source, $k$-eigenvalue, and neutron noise problems, using both shared and distributed memory parallelism. Shared memory parallelism is implemented with OpenMP, while the distributed memory parallelism is implemented using MPI. Different transport methods such as surface-tracking, delta-tracking, and negative-weighted delta-tracking are also available. All of the outlined cancellation algorithms have been implemented in MGMC, which was used to run the simulations presented in the next section. MGMC has been make publicly available as free software under the CeCILL v2.1 license \cite{MGMC}.
        
    \section{Simulation Results}\label{sec:results}
    
        For our numerical simulations, in this Section we will make use of the modified C5G7 benchmark which we introduced in our previous work \cite{Belanger2021}. The C5G7 is a multi-group neutron transport benchmark which comes from the nuclear reactor physics community, for the purpose of validating different codes \cite{C5G7}. Our modified version makes use of square profile fuel pins with side lengths of $\SI{0.756}{\centi\meter}$, in lieu of cylindrical pins of radius $\SI{0.54}{\centi\meter}$ as proposed in the original specifications. This modification allows a regular $170\times 170\times 765$ mesh to be imposed on top of the geometry over the fuel assemblies to act as cancellation regions, and guaranteed that each cancellation region contained a unique material. For continuity, we make use of the same cancellation mesh. For transport, we again use the negative-weighted delta-tracking variant proposed by Carter et al.\ \cite{Carter1972}, and identical sampling cross sections to the previous study: the sampling cross section for the first group is 90\% of the majorant cross section, while all other sampling cross sections were taken to be the majorant. This means that the sampling cross section underestimates the total cross section in the first energy group for all fuel pins in the problem. Whenever a virtual collision occurs for a particle in the first energy group, inside a fuel pin, its weight will then change sign. Once a particle leaves the first group, it is impossible for the sign to change at a collision (although signs can possibly change during cancellation).
        Virtual collisions lead to the presence of negative weights in the system, and we have shown that weight cancellation is mandatory for $k$-eigenvalue power iteration problem to converge when using negative-weighted delta-tracking \cite{Belanger2021}. All simulations were initiated with $10^6$ particles, and ran for 2500 generations, with the first 500 being discarded to allow for source convergence.
        
        As we have shown in Ref.~\citenum{Belanger2021}, the total weight of all the fission particles between two generations increases without bound if weight cancellation is not applied. This increase in total weight is accompanied by an increase in the number of particles and large statistical fluctuations in estimated quantities, making it nearly impossible to estimate the multiplication factor and static flux for the system. The effect of cancellation is to limit the growth of $W_\text{tot}$ to a saturation value; the more efficient cancellation is, the lower the saturation value will be. Thus, we have chosen to assess the efficiency of cancellation by using the saturation value of $W_\text{tot}$, which is calculated immediately after applying the cancellation procedure. Note that $W_\text{tot}$ has a lower theoretical limit of $W_\text{net}$, which is kept constant by normalizing all particle weights between generations \cite{Belanger2021}.

        \subsection{Comparison of Optimization Strategies}\label{opt_comparison}
\begin{figure}
                \centering
               \includegraphics[width=\columnwidth]{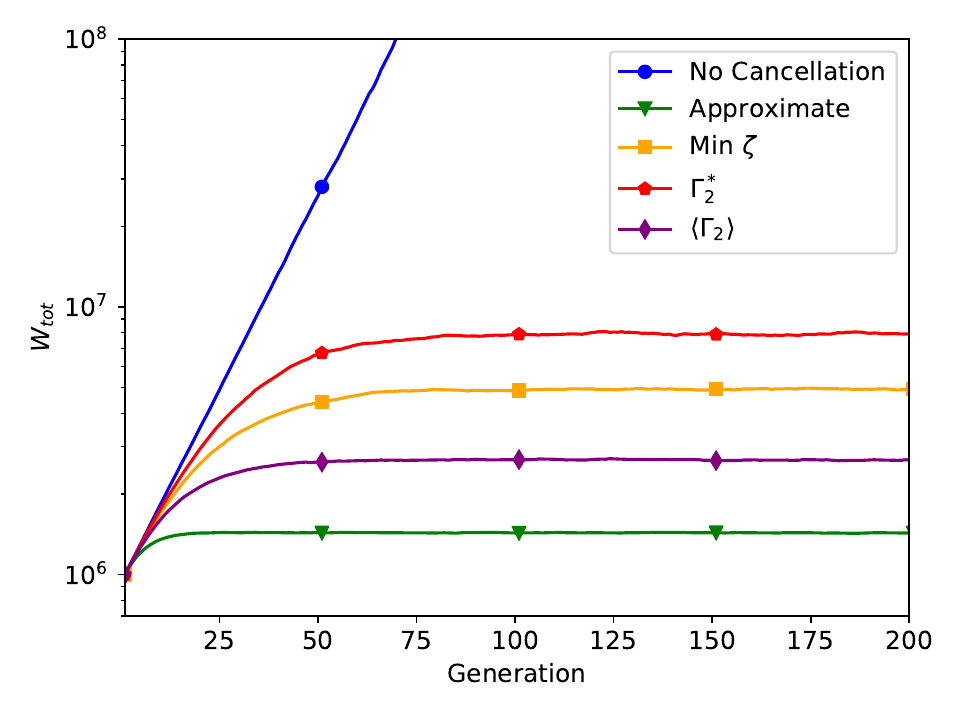}
                \caption{Behavior of $W_\text{tot}$ as a function of generation, for different cancellation methods, including no cancellation and approximate cancellation.}
                \label{fig:opt_comparison}
            \end{figure}

            To determine which method of choosing $\beta_k$ leads to the most efficient cancellation of positive and negative weights, the optimization techniques described in Sec.~\ref{optimization_replacement} and Sec.~\ref{optimization_average} were compared against the original implementation using the minimum value of the fission density within the region. Both optimization options utilized $N_s=100$ samples for estimating $\expval{\zeta_k}$ and $\expval{1/\zeta_k}$. The values of $W_\text{tot}$ are plotted against the number of generations in Figure~\ref{fig:opt_comparison}. For comparison, curves corresponding to no cancellation and approximate cancellation have also been presented. Approximate cancellation imposes a mesh on top of the geometry, and sorts fission particle into this mesh. The average weight of all particles in each mesh element can then be calculated and assigned to the particles \cite{Zhang2016,BelangerMC2021}. This method is quite efficient, but is not exact, and imposes a bias on the fission source and on the eigenvalue (though the bias can be made arbitrarily small by using a sufficiently fine mesh).
            
            If no cancellation technique is used, the total weight increases exponentially, without bound. This phenomenon is expected when using negative-weighted delta-tracking with $k$-eigenvalue power iteration, as described previously \cite{Belanger2021}. When taking $\beta_k = \min_\RR(\zeta_k)$, an asymptotic value of $W_\text{tot}\approx 4.9\cdot 10^6$ was seen. The most efficient method of determining $\beta_k$ was that obtained from optimizing $\expval{\Gamma_2}$ in Eq.~\eqref{eq:opt2_beta_j}, resulting in $W_\text{tot}\approx 2.7\cdot 10^6$, almost half the amount of total weight obtained with the minimum strategy. Calculating $\beta_k$ from Eq.~\eqref{eq:opt_beta_j} for the case of replacing $\zeta_k$ with $\expval{\zeta_k}$ is less efficient than using the minimum value of $\zeta_k$ within the cancellation region, resulting in $W_\text{tot}\approx 8.0\cdot 10^6$. It is not known why this approximation does not perform as well as using the minimum of $\zeta_k$, and this intriguing question calls for future investigations.
            
            Approximate cancellation yielded the lowest total weight (and therefore the highest cancellation efficiency), with $W_\text{tot}\approx 1.4\cdot 10^6$, but is not an exact approach. Currently, we do not know of any way to estimate, or to put a limit on the bias imposed by this method without running several realizations, each with a different mesh size.
        
        \subsection{Strategies for Evaluating the Average Fission Emission Densities}\label{sobol_v_prng}
            
\begin{figure}
                \centering
                \includegraphics[width=\columnwidth]{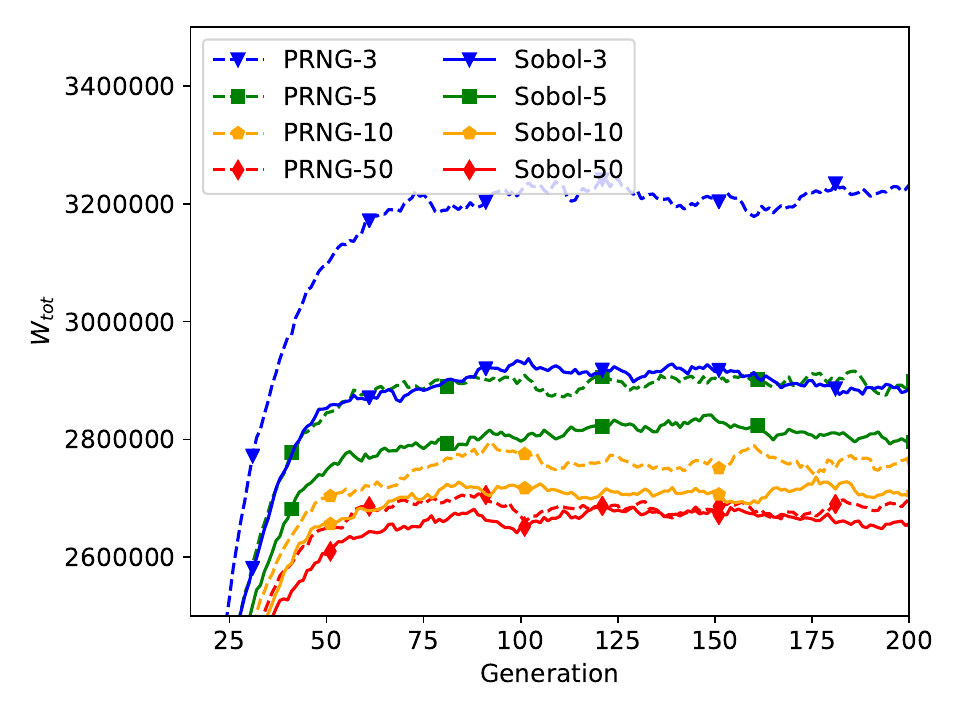}
                \caption{Behaviour of $W_\text{tot}$ as a function of generation, for different cancellation parameters. For all curves, $\beta_k$ is determined from the optimization of $\expval{\Gamma_2}$, Eq.~\eqref{eq:opt2_beta_j}. The estimated values of $\expval{\zeta_k}$ and $\expval{1/\zeta_k}$ are determined with a varying number of points, using either a pseudo-random number generator (PRNG) or a Sobol' sequence.}
                \label{fig:prng_v_sobol}
            \end{figure}

            The analysis in Sec.~\ref{opt_comparison} shows that the optimal choice for determining $\beta_k$ for the C5G7 benchmark is Eq.~\eqref{eq:opt2_beta_j}, from the optimization of $\expval{\Gamma_2}$. We now consider the optimal strategy for estimating the requisite values of $\expval{\zeta_k}$ and $\expval{1/\zeta_k}$ for each particle. Figure~\ref{fig:prng_v_sobol} depicts the behavior of $W_\text{tot}$ where points $\tilde{P}_i$ (from Eqs.~\eqref{eq:exp_fk} and \eqref{eq:exp_1_fk}) are sampled with either a pseudo-random number generator (PRNG) or a Sobol' sequence.
            
            First of all, the spread among the different strategies for estimating $\expval{\zeta_k}$ and $\expval{1/\zeta_k}$ is much smaller than the spread among the different minimization strategies of Fig.~\ref{fig:opt_comparison}. 
            It is observed that in general, when $N_s < 50$, using Sobol' points leads to more efficient weight cancellation. This effect is most apparent for $N_s = 3$, where the Sobol' sequence leads to approximately 9.2\% less total weight being transported, compared to the PRNG estimation strategy. The increased efficiency observed in the Sobol' points diminishes however with increasing $N_s$. Sobol' estimation gives a 2.8\% improvement for $N_s=5$, 1.8\% improvement for $N_s=10$, and only a 0.5\% improvement for $N_s=50$. This would indicate that the estimated values for $\expval{\zeta_k}$ and $\expval{1/\zeta_k}$ start to become independent of the evaluation strategy at around $N_s=50$.
            
            In addition to achieving more weight cancellation, the Sobol' points also have the added benefit of being slightly easier to compute, as the quasi-random numbers used the compute the points can be tabulated in advance, and written in the code. All that is then needed is a table lookup to get a Sobol' value, whereas several mathematical operations must be performed to calculate each value generated from a PRNG. However, a drawback with the use of Sobol' points is that one does not necessarily know in advance how many points will be needed, when considering heterogeneous cancellation regions.
        
        \subsection{Demonstration of Heterogeneous Cancellation Regions on the C5G7 Benchmark}
            
            We also tested the rejection-based sampling technique described in Section~\ref{heterogeneous_regions}, for performing regional cancellation in cuboid regions which contain multiple materials. Instead of using our modified version of the C5G7 benchmark, we have opted to use the original version with cylindrical fuel pins \cite{C5G7}, in combination with the same $170\times 170\times 765$ mesh as used in our previous simulations. A reference calculation was performed using standard delta-tracking and obtained a multiplication factor of $\keff = 1.18383 \pm 0.00003$, which is in agreement with the reference solution for the 3D version of the benchmark \cite{C5G7}. Cancellation used the method for calculating $\beta$ proposed in Sec.~\ref{optimization_average}, with $N_s=10$ samples being used to estimate $\expval{\zeta_k}$ and $\expval{1/\zeta_k}$.
            
            \begin{figure}
                \centering
                \includegraphics[width=\columnwidth]{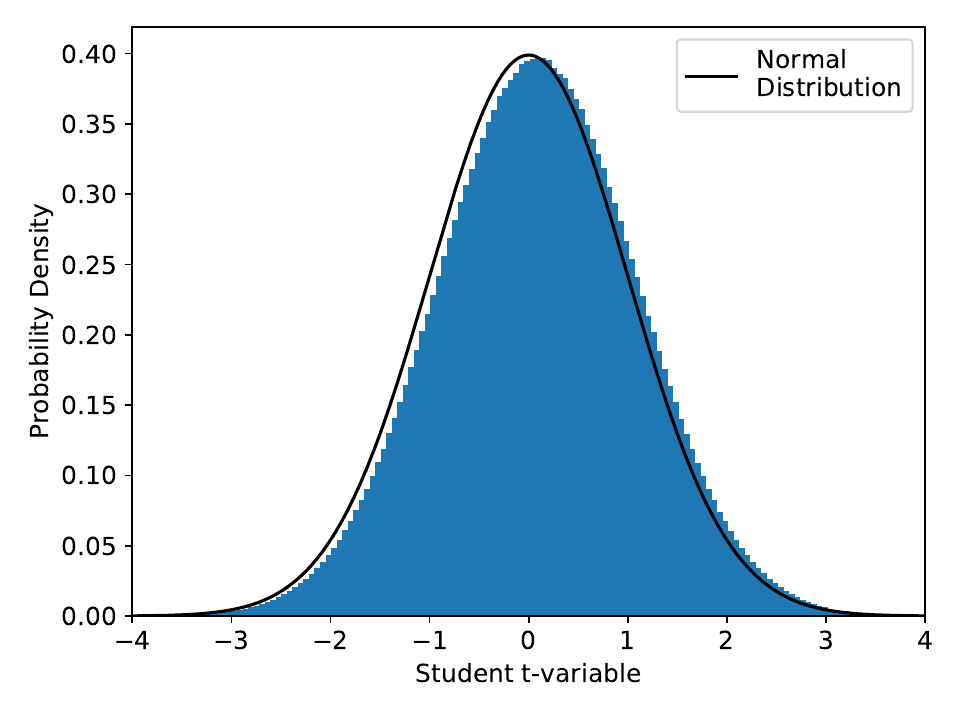}
                \caption{Histogram distribution of the Student $t$-variable, comparing the flux computed from delta-tracking to the flux computed with negative weighted delta-tracking. A normal distribution is plotted on top of the histogram for a reference.}
                \label{fig:c5g7_t_variable}
            \end{figure}
            
            \begin{figure}
                \centering
                \includegraphics[width=\columnwidth]{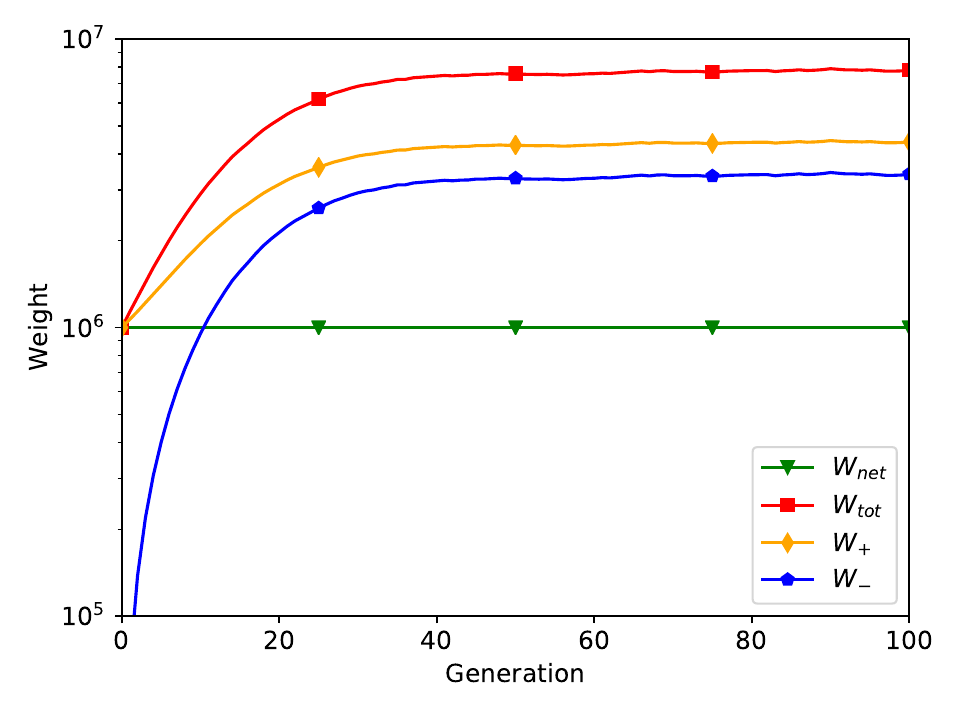}
                \caption{Plots of the positive, negative, net, and total weights as a function of generation, for the original C5G7 benchmark. Exact regional cancellation with heterogeneous regions was used to perform the simulation.}
                \label{fig:c5g7_wgt}
            \end{figure}
            
            When running the same simulation with negative-weighted delta-tracking, using the same sampling cross sections as before, and exact cancellation, a multiplication factor of $\keff = 1.18382 \pm 0.00009$ was obtained, which is in agreement with the delta-tracking value. A comparison of the two estimations of the flux were also made, looking at the Student $t$-variable, which is defined as
            \begin{equation}
            t_i=\frac{\varphi_{i,A}-\varphi_{i,B}}{\sqrt{\sigma^2_{i,A}+\sigma^2_{i,B}}}
            \text,
            \end{equation}
            where $\varphi_{i,x}$ is the average value of the flux in the $i$-th bin for calculation $x$, and $\sigma_{i,x}$ is its standard error. $A$ corresponds to the results from the delta-tracking simulation without weight cancellation, and $B$ corresponds to the results from the negative-weighted delta-tracking simulation with cancellation. For independent, normally distributed variables with a large number of degrees of freedom, the distribution of the $t$-variable should approach a normal distribution. A plot of the empirical distribution of the $t$-variable is provided in Figure~\ref{fig:c5g7_t_variable}. We excluded from the comparison all the bins where the flux was estimated to be zero, or where the relative standard error was greater than 20\%. This was done in an effort to ensure that each bin was approximately normally distributed, for the Student $t$-variable distribution assumptions to be reasonable. It is clear from Figure~\ref{fig:c5g7_t_variable} that the two flux estimates are in good agreement. The $t$-variable distribution is not perfectly normal, which is to be expected, as there are correlations between the scores in different flux bins. In general, this is a very strong indication that our cancellation method has not imposed any bias on the fission source, and that the method is still exact when applied to heterogeneous cancellation regions.
            
            The behavior of the total weight is shown in Fig.~\ref{fig:c5g7_wgt}. While $W_\text{tot}$ is larger than in the case of square fuel pins with homogeneous cancellation region presented in Fig.~\ref{fig:opt_comparison}, the behavior is in general similar. A large $W_\text{tot}$ indicates that there are more negative particles, which will increase the variance in scores. This is indeed the case, as our previous work obtained an uncertainty for $\keff$ of only $5\times 10^{-5}$ (albeit for a slightly different problem) \cite{Belanger2021}, while an uncertainty of $9\times 10^{-5}$ was obtained for this problem.
        
    \section{Conclusions}\label{sec:conclusions}
        
        This work has leveraged the integral form of the Boltzmann transport equation to provide a more in-depth mathematical analysis of the exact regional weight cancellation technique, considerably expanding on previous works on the subject \cite{Booth2010,Belanger2021}. Not only has this formal approach given a much better understanding as to the mechanics of regional cancellation in simplified isotropic multi-group problems, but it has illuminated the non-trivial path to performing exact cancellation in more complex problems, where scattering is anisotropic, and the fission spectrum may depend on the incident energy of a particle. The analysis highlights the fact that the implementation of exact regional cancellation is rather straightforward in a multi-group Monte Carlo code, but will be more difficult in a continuous-energy code, as one will need access to conditional scattering distributions in the laboratory frame, which are not always available for all reactions. The implementation of exact regional cancellation in a continuous-energy code is thus a subject which will require further research.
        
        Additionally, a strategy to determine an optimal value of the cancellation parameter $\beta$ for each particle undergoing cancellation has been conceived. For each particle $k$ within a cancellation region, its optimal cancellation parameter $\beta_k$ can be computed if both the average value of the fission emission density in the region and the average value of the inverse of the fission emission density in the region are known. Our previous implementation required that cancellation regions be homogeneous and cubical, which restricted its applicability to simple problems. Thanks to the improvements in the optimization technique proposed in this work, both requirements have been relaxed.
        
        On a modified version of the C5G7 benchmark, our technique to optimize weight cancellation was demonstrated to reduce the total weight in the simulation by approximately 45\%, when compared to using the minimum value of the fission emission density in the region for $\beta$, as previously suggested in the literature. In order to estimate the average fission emission density and the average inverse of the fission emission density, a sampling approach has been proposed, where the averages are estimated using the values of the fission emission density at pseudo-random or quasi-random points. It was demonstrated that the quasi-random Sobol' sequence requires slightly fewer points than the pseudo-random sequence to reach the asymptotic limit of the optimized cancellation algorithm. As a comparison, the use of 3 Sobol' points had very similar performance to the use of 5 pseudo-random points for the modified version of the C5G7 benchmark examined here. However, such results are likely to be highly problem-dependent, and more systems should be analyzed to ascertain what sort of performance improvements could be expected in general. We also tested the use of heterogeneous cancellation regions on the original C5G7 benchmark, with cylindrical fuel pins. No bias was observed in the resulting fundamental eigenvalue, or flux tally.
        
    \appendix
    
    \section{Optimization of $\Gamma_2^*$}\label{ap:opt_derivation}
        We remind the reader of the definition of $\Gamma_2^*$:
        \begin{equation}
            \Gamma^*_{2} = \sum_{k=1}^{N}\bigg(\frac{\expval{\zeta_k} - \beta_k}{\expval{\zeta_k}}w_k\bigg)^2 + \bigg(\sum_{k=1}^{N}\frac{\beta_k}{\expval{\zeta_k}}w_k\bigg)^2.
        \end{equation}
        We optimize $\Gamma_2^*$ simultaneously for all particles by differentiating with respect to $\beta_j$, and setting the partial derivative equal to zero:
        \begin{equation}
            \frac{\partial\Gamma_2^*}{\partial\beta_j} =
            -2\frac{\expval{\zeta_j} - \beta_j}{\expval{\zeta_j}^2}w^2_j +
            2\frac{w_j}{\expval{\zeta_j}}
            \sum_{k=1}^{N}\frac{\beta_k w_k}{\expval{\zeta_k}}=0.
        \end{equation}
        This may be simplified to
        \begin{equation}
            -w_j + \frac{\beta_j w_j}{\expval{\zeta_j}} + \sum_{k=1}^{N}\frac{\beta_k w_k}{\expval{\zeta_k}} = 0.
            \label{eq:opt_with_sum}
        \end{equation}
        On the left-hand-side, the second term matches the argument of the sum
        in the third term. Summing over index $j$, we see that
        \begin{equation}
            -W + \sum_{j=1}^{N}\frac{\beta_j w_j}{\expval{\zeta_j}} +
            N\sum_{k=1}^{N}\frac{\beta_k w_k}{\expval{\zeta_k}} = 0.
        \end{equation}
        Here, we used the definition provided in Eq.~\eqref{eq:W}.
        This allows us to isolate the sum
        \begin{equation}
            \sum_{k=1}^{N}\frac{\beta_k w_k}{\expval{\zeta_k}} = S^* = \frac{W}{N+1}.
        \end{equation}
        Applying this substitution to Eq.~\eqref{eq:opt_with_sum} while also using Eqs.~\eqref{eq:S_tilde} and \eqref{eq:W}, we find that
        the optimized value of $\beta_j$ is
        \begin{equation}
            \beta_j = \expval{\zeta_j} \left(1  - \frac{W}{(N+1)w_j} \right) = \expval{\zeta_j} \left(1  - \frac{S^*}{w_j} \right).
        \end{equation}
    
    \section{Optimization of $\expval{\Gamma_2}$}\label{ap:opt2_derivation}
        Substituting Eq.~\eqref{eq:L2_norm} into Eq.~\eqref{eq:avg_l2}, and partially expanding the squared terms, we see that
        \begin{widetext}
        \begin{align}
            \expval{\Gamma_2} &=
            \frac{\displaystyle\int_\mathcal{R}\bigg[\sum_{k=1}^{N}
            \bigg(1 - \frac{2\beta_k}{\zeta_k} + \frac{\beta^2_k}{\zeta^2_k}\bigg)w^2_k + 
            \sum_{k=1}^{N}\sum_{l=1}^{N}
            \frac{\beta_k}{\zeta_k}\frac{\beta_l}{\zeta_l}w_kw_l\bigg]
            \prod_{m=1}^{N}\zeta_m\dd P_m}
            {\displaystyle\int_\mathcal{R}\prod_{n=1}^{N}\zeta_n\dd P_n}\\
            &=
            \frac{\displaystyle\int_\mathcal{R}\bigg[\sum_{k=1}^{N}
            \bigg(1 - \frac{2\beta_k}{\zeta_k} + \frac{\beta^2_k}{\zeta^2_k}\bigg)w^2_k + 
            \sum_{k=1}^{N}\sum_{\substack{l=1 \\ l\ne k}}^{N}
            \frac{\beta_k}{\zeta_k}\frac{\beta_l}{\zeta_l}w_kw_l + \sum_{k=1}^{N}\frac{\beta_k^2}{\zeta_k^2}w_k^2\bigg]
            \prod_{m=1}^{N}\zeta_m\dd P_m}
            {\displaystyle\prod_{n=1}^{N}\expval{\zeta_n}}\\
            &=
            \sum_{k=1}^{N}
            \bigg(1 - \frac{2\beta_k}{\expval{\zeta_k}} + \frac{\beta^2_k}{\expval{\zeta_k}}\expval{\frac{1}{\zeta_k}}\bigg)w^2_k + 
            \sum_{k=1}^{N}\sum_{\substack{l=1 \\ l\ne k}}^{N}
            \frac{\beta_k}{\expval{\zeta_k}}\frac{\beta_l}{\expval{\zeta_l}}w_kw_l + \sum_{k=1}^{N}\frac{\beta_k^2}{\expval{\zeta_k}}\expval{\frac{1}{\zeta_k}}w_k^2\\
            &=
            \sum_{k=1}^{N}
            \bigg(w^2_k - \frac{2\beta_kw^2_k}{\expval{\zeta_k}}\bigg) +
            \sum_{k=1}^{N}\frac{\beta^2_kw^2_k}{\expval{\zeta_k}}\bigg(2\expval{\frac{1}{\zeta_k}} - \frac{1}{\expval{\zeta_k}}\bigg) + 
            \sum_{k=1}^{N}\sum_{l=1}^{N}
            \frac{\beta_k}{\expval{\zeta_k}}\frac{\beta_l}{\expval{\zeta_l}}w_kw_l.
            \label{eq:pre_ck}
        \end{align}
        It is convenient to use the constant $c_k$, defined by Eq.~\eqref{eq:ck}, which may be substituted into Eq.~\eqref{eq:pre_ck} to produce
        \begin{equation}
            \expval{\Gamma_2} = \sum_{k=1}^{N}
            \bigg(w^2_k - \frac{2\beta_kw^2_k}{\expval{\zeta_k}}\bigg) +
            \sum_{k=1}^{N}\frac{\beta^2_kw^2_k}{c_k\expval{\zeta_k}^2} + 
            \sum_{k=1}^{N}\sum_{l=1}^{N}
            \frac{\beta_k}{\expval{\zeta_k}}\frac{\beta_l}{\expval{\zeta_l}}w_kw_l.
            \label{eq:expval_G2_final}
        \end{equation}
        \end{widetext}
        
        Now that all of the integrals have been simplified, we are left with $\expval{\Gamma_2}$ as a function of $\beta_k$, $\expval{\zeta_k}$, and $\expval{\frac{1}{\zeta_k}}$ $\forall k = 1, \dots, N$. We now optimize $\expval{\Gamma_2}$ with respect to the cancellation parameter $\beta_j$ by solving for
        \begin{equation}
            \frac{\partial\expval{\Gamma_2}}{\partial\beta_j} = 0.
        \end{equation}
        From Eq.~\eqref{eq:expval_G2_final}, one may then proceed by solving
        \begin{equation}
            \frac{\partial\expval{\Gamma_2}}{\partial\beta_j} =
            - \frac{2w_j^2}{\expval{\zeta_j}}
            + \frac{2\beta_j w_j^2}{c_j\expval{\zeta_j}^2}
            + \frac{2w_j}{\expval{\zeta_j}}\sum_{k=1}^{N}\frac{\beta_k w_k}{\expval{\zeta_k}} = 0.
        \end{equation}
        Upon a division by $2w_j/\expval{\zeta_j}$ on both sides,
        we are left with
        \begin{equation}
            -w_j + \frac{\beta_j w_j}{c_j\expval{\zeta_j}} + \sum_{k=1}^{N}\frac{\beta_k w_k}{\expval{\zeta_k}} = 0.
            \label{eq:opt2_with_sum}
        \end{equation}
        It is possible to isolate the sum in the third term on the left-hand-side by multiplying by $c_j$, and then summing over $j$:
        \begin{equation}
            -\sum_{j=1}^{N}c_j w_j + \sum_{j=1}^{N}\frac{\beta_j w_j}{\expval{\zeta_k}} + \sum_{j=1}^{N}c_j\sum_{k=1}^{N}\frac{\beta_k w_k}{\expval{\zeta_k}} = 0.
            \label{eq:opt2_pre_S}
        \end{equation}
        We will now define
        \begin{equation}
            S = \sum_{k=1}^{N} \frac{\beta_k w_k}{\expval{\zeta_k}},
            \label{eq:S_sub}
        \end{equation}
        and substitute Eq.~\eqref{eq:S_sub} into Eq.~\eqref{eq:opt2_pre_S}, allowing one to solve for $S$. Doing so, one may obtain the result provided by Eq.~\eqref{eq:S}. Now that the summation term, $S$, can be computed without knowledge of $\beta_j$, we may substitute Eq.~\eqref{eq:S} and Eq.~\eqref{eq:S_sub} into Eq.~\eqref{eq:opt2_with_sum}, and solve for $\beta_j$, producing
        \begin{equation}
           \beta_j = \expval{\zeta_j}c_j\bigg(1-\frac{S}{w_j}\bigg).
        \end{equation}
        
%

\end{document}